\documentclass[%
 reprint,
superscriptaddress,
showpacs,preprintnumbers,
 amsmath,amssymb,
 aps,
]{revtex4-1}

\usepackage{graphicx}
\usepackage{dcolumn}
\usepackage{bm}

\begin{document}
\renewcommand{\textfraction}{0.00000000001}
\renewcommand{\floatpagefraction}{1.0}

\preprint{APS/123-QED}

\title{Photoproduction of {\boldmath{$\pi^{0}$}} Mesons off Protons and Neutrons\\ 
in the Second and Third Nucleon Resonance Region}
\author{M.~Dieterle}
\affiliation{Department of Physics, University of Basel, Ch-4056 Basel, Switzerland}
\author{D.~Werthm\"uller}
\affiliation{Department of Physics, University of Basel, Ch-4056 Basel, Switzerland}
\affiliation{SUPA School of Physics and Astronomy, University of Glasgow, Glasgow, G12 8QQ, UK}
\author{S.~Abt}
\affiliation{Department of Physics, University of Basel, Ch-4056 Basel, Switzerland}
\author{F.~Afzal}
\affiliation{Helmholtz-Institut f\"ur Strahlen- und Kernphysik, University Bonn, D-53115 Bonn, Germany}
\author{P.~Aguar Bartolome}
\affiliation{Institut f\"ur Kernphysik, University of Mainz, D-55099 Mainz, Germany}
\author{Z.~Ahmed}
\affiliation{University of Regina, Regina, SK S4S-0A2 Canada}
\author{J.~Ahrens}
\affiliation{Institut f\"ur Kernphysik, University of Mainz, D-55099 Mainz, Germany}
\author{J.R.M.~Annand}
\affiliation{SUPA School of Physics and Astronomy, University of Glasgow, Glasgow, G12 8QQ, UK}
\author{H.J.~Arends}
\affiliation{Institut f\"ur Kernphysik, University of Mainz, D-55099 Mainz, Germany}
\author{M.~Bashkanov}
\affiliation{SUPA School of Physics, University of Edinburgh, Edinburgh EEH9 3JZ, UK}
\author{R.~Beck}
\affiliation{Helmholtz-Institut f\"ur Strahlen- und Kernphysik, University Bonn, D-53115 Bonn, Germany}
\author{M.~Biroth}
\affiliation{Institut f\"ur Kernphysik, University of Mainz, D-55099 Mainz, Germany}
\author{N.~Borisov}
\affiliation{Joint Institute for Nuclear Research, 141980 Dubna, Russia}  
\author{A.~Braghieri}
\affiliation{INFN Sezione di Pavia, I-27100 Pavia, Pavia, Italy}
\author{W.J.~Briscoe}
\affiliation{Center for Nuclear Studies, The George Washington University, Washington, DC 20052, USA}
\author{S.~Cherepnya}
\affiliation{Lebedev Physical Institute, RU-119991 Moscow, Russia}
\author{F.~Cividini}
\affiliation{Institut f\"ur Kernphysik, University of Mainz, D-55099 Mainz, Germany}
\author{C.~Collicott}
\affiliation{Department of Astronomy and Physics, Saint Mary's University, E4L1E6 Halifax, Canada}
\author{S.~Costanza}\altaffiliation{Also at: Dipartimento di Fisica, Universit\`a di Pavia, I-27100 Pavia, Italy}
\affiliation{INFN Sezione di Pavia, I-27100 Pavia, Pavia, Italy}
\author{A.~Denig}
\affiliation{Institut f\"ur Kernphysik, University of Mainz, D-55099 Mainz, Germany}
\author{E.J.~Downie}
\affiliation{Center for Nuclear Studies, The George Washington University, Washington, DC 20052, USA}
\author{P.~Drexler}
\affiliation{Institut f\"ur Kernphysik, University of Mainz, D-55099 Mainz, Germany}
\affiliation{II. Physikalisches Institut, University of Giessen, D-35392 Giessen, Germany}
\author{L.V.~Fil'kov}
\affiliation{Lebedev Physical Institute, RU-119991 Moscow, Russia}
\author{S.~Garni}
\affiliation{Department of Physics, University of Basel, Ch-4056 Basel, Switzerland}  
\author{D.I.~Glazier}
\affiliation{SUPA School of Physics and Astronomy, University of Glasgow, Glasgow, G12 8QQ, UK}
\affiliation{SUPA School of Physics, University of Edinburgh, Edinburgh EEH9 3JZ, UK}
\author{I.~Gorodnov}
\affiliation{Joint Institute for Nuclear Research, 141980 Dubna, Russia}
\author{W.~Gradl}
\affiliation{Institut f\"ur Kernphysik, University of Mainz, D-55099 Mainz, Germany}
\author{M.~G{\"u}nther}
\affiliation{Department of Physics, University of Basel, Ch-4056 Basel, Switzerland}   
\author{D.~Gurevich}
\affiliation{Institute for Nuclear Research, RU-125047 Moscow, Russia}
\author{L. Heijkenskj{\"o}ld}
\affiliation{Institut f\"ur Kernphysik, University of Mainz, D-55099 Mainz, Germany}
\author{D.~Hornidge}
\affiliation{Mount Allison University, Sackville, New Brunswick E4L1E6, Canada}
\author{G.M.~Huber}
\affiliation{University of Regina, Regina, SK S4S-0A2 Canada}
\author{A.~K{\"a}ser}
\affiliation{Department of Physics, University of Basel, Ch-4056 Basel, Switzerland}   
\author{V.L.~Kashevarov}
\affiliation{Institut f\"ur Kernphysik, University of Mainz, D-55099 Mainz, Germany}
\affiliation{Joint Institute for Nuclear Research, 141980 Dubna, Russia}
\author{S.~Kay}
\affiliation{SUPA School of Physics, University of Edinburgh, Edinburgh EEH9 3JZ, UK}
\author{I.~Keshelashvili}\altaffiliation{Present adaress: Institut f\"ur Kernphysik, FZ J\"ulich, 52425 J\"ulich, Germany}
\affiliation{Department of Physics, University of Basel, Ch-4056 Basel, Switzerland}
\author{R.~Kondratiev}
\affiliation{Institute for Nuclear Research, RU-125047 Moscow, Russia}
\author{M.~Korolija}
\affiliation{Rudjer Boskovic Institute, HR-10000 Zagreb, Croatia}
\author{B.~Krusche}\email[]{Corresponding author: email bernd.krusche@unibas.ch}
\affiliation{Department of Physics, University of Basel, Ch-4056 Basel, Switzerland}
\author{A.~Lazarev}
\affiliation{Joint Institute for Nuclear Research, 141980 Dubna, Russia}  
\author{V.~Lisin}
\affiliation{Institute for Nuclear Research, RU-125047 Moscow, Russia}
\author{K.~Livingston}
\affiliation{SUPA School of Physics and Astronomy, University of Glasgow, Glasgow, G12 8QQ, UK}
\author{S.~Lutterer}
\affiliation{Department of Physics, University of Basel, Ch-4056 Basel, Switzerland}
\author{I.J.D.~MacGregor}
\affiliation{SUPA School of Physics and Astronomy, University of Glasgow, Glasgow, G12 8QQ, UK}
\author{D.M.~Manley}
\affiliation{Kent State University, Kent, Ohio 44242, USA}
\author{P.P.~Martel}
\affiliation{Institut f\"ur Kernphysik, University of Mainz, D-55099 Mainz, Germany}
\affiliation{Mount Allison University, Sackville, New Brunswick E4L3B5, Canada}
\author{J.C.~McGeorge}
\affiliation{SUPA School of Physics and Astronomy, University of Glasgow, Glasgow, G12 8QQ, UK}
\author{V.~Metag}
\affiliation{II. Physikalisches Institut, University of Giessen, D-35392 Giessen, Germany}
\author{D.G.~Middleton}
\affiliation{Mount Allison University, Sackville, New Brunswick E4L3B5, Canada}
\author{R.~Miskimen}
\affiliation{University of Massachusetts, Amherst, Massachusetts 01003, USA}
\author{E.~Mornacchi}
\affiliation{Institut f\"ur Kernphysik, University of Mainz, D-55099 Mainz, Germany}
\author{A.~Mushkarenkov}
\affiliation{INFN Sezione di Pavia, I-27100 Pavia, Pavia, Italy}  
\affiliation{University of Massachusetts, Amherst, Massachusetts 01003, USA}
\author{A.~Neganov}
\affiliation{Joint Institute for Nuclear Research, 141980 Dubna, Russia}  
\author{A.~Neiser}
\affiliation{Institut f\"ur Kernphysik, University of Mainz, D-55099 Mainz, Germany}
\author{M.~Oberle}
\affiliation{Department of Physics, University of Basel, Ch-4056 Basel, Switzerland}  
\author{M.~Ostrick}
\affiliation{Institut f\"ur Kernphysik, University of Mainz, D-55099 Mainz, Germany}
\author{P.B.~Otte}
\affiliation{Institut f\"ur Kernphysik, University of Mainz, D-55099 Mainz, Germany}
\author{B.~Oussena}
\affiliation{Institut f\"ur Kernphysik, University of Mainz, D-55099 Mainz, Germany}
\affiliation{Center for Nuclear Studies, The George Washington University, Washington, DC 20052, USA}
\author{D.~Paudyal}
\affiliation{University of Regina, Regina, SK S4S-0A2 Canada}
\author{P.~Pedroni}
\affiliation{INFN Sezione di Pavia, I-27100 Pavia, Pavia, Italy}
\author{A.~Polonski}
\affiliation{Institute for Nuclear Research, RU-125047 Moscow, Russia}
\author{S.N.~Prakhov}
\affiliation{University of California Los Angeles, Los Angeles, California 90095-1547, USA}
\author{G.~Ron}
\affiliation{Racah Institute of Physics, Hebrew University of Jerusalem, Jerusalem 91904, Israel}
\author{T.~Rostomyan}\altaffiliation{Present address: Department of Physics and Astronomy, Rutgers University,
Piscataway, New Jersey, 08854-8019}
\affiliation{Department of Physics, University of Basel, Ch-4056 Basel, Switzerland}
\author{A.~Sarty}
\affiliation{Department of Astronomy and Physics, Saint Mary's University, E4L1E6 Halifax, Canada}
\author{C.~Sfienti}
\affiliation{Institut f\"ur Kernphysik, University of Mainz, D-55099 Mainz, Germany}
\author{V.~Sokhoyan}
\affiliation{Institut f\"ur Kernphysik, University of Mainz, D-55099 Mainz, Germany}
\author{K.~Spieker}
\affiliation{Helmholtz-Institut f\"ur Strahlen- und Kernphysik, University Bonn, D-53115 Bonn, Germany}
\author{O.~Steffen}
\affiliation{Institut f\"ur Kernphysik, University of Mainz, D-55099 Mainz, Germany}
\author{I.I.~Strakovsky}
\affiliation{Center for Nuclear Studies, The George Washington University, Washington, DC 20052, USA}
\author{T.~Strub}
\affiliation{Department of Physics, University of Basel, Ch-4056 Basel, Switzerland}
\author{I.~Supek}
\affiliation{Rudjer Boskovic Institute, HR-10000 Zagreb, Croatia}
\author{A.~Thiel}
\affiliation{Helmholtz-Institut f\"ur Strahlen- und Kernphysik, University Bonn, D-53115 Bonn, Germany}
\author{M.~Thiel}
\affiliation{Institut f\"ur Kernphysik, University of Mainz, D-55099 Mainz, Germany}
\author{A.~Thomas}
\affiliation{Institut f\"ur Kernphysik, University of Mainz, D-55099 Mainz, Germany}
\author{M.~Unverzagt}
\affiliation{Institut f\"ur Kernphysik, University of Mainz, D-55099 Mainz, Germany}
\author{Yu.A.~Usov}
\affiliation{Joint Institute for Nuclear Research, 141980 Dubna, Russia}  
\author{S.~Wagner}
\affiliation{Institut f\"ur Kernphysik, University of Mainz, D-55099 Mainz, Germany}
\author{N.K.~Walford}
\affiliation{Department of Physics, University of Basel, Ch-4056 Basel, Switzerland}
\author{D.P.~Watts}
\affiliation{SUPA School of Physics, University of Edinburgh, Edinburgh EEH9 3JZ, UK}
\author{J.~Wettig}
\affiliation{Institut f\"ur Kernphysik, University of Mainz, D-55099 Mainz, Germany}
\author{L.~Witthauer}
\affiliation{Department of Physics, University of Basel, Ch-4056 Basel, Switzerland}
\author{M.~Wolfes}
\affiliation{Institut f\"ur Kernphysik, University of Mainz, D-55099 Mainz, Germany}
\author{L.A.~Zana}
\affiliation{SUPA School of Physics, University of Edinburgh, Edinburgh EEH9 3JZ, UK}
\collaboration{A2 Collaboration}

\date{\today}

\begin{abstract}
\begin{description}
\item[Background] 
{Photoproduction of mesons off quasi-free nucleons bound in the deuteron allows to study
the electromagnetic excitation spectrum of the neutron and the isospin structure
of the excitation of nucleon resonances. The database for such reactions is much more 
sparse than for free proton targets.}
\item[Purpose] 
{Study experimentally single $\pi^0$ photoproduction off quasi-free nucleons
from the deuteron.
Investigate nuclear effects by a comparison of the results for free protons and quasi-free protons. 
Use the quasi-free neutron data (corrected for nuclear effects) to test the predictions of reaction 
models and partial wave analysis (PWA) for $\gamma n\rightarrow n\pi^0$ derived from the analysis 
of the other isospin channels.}
\item[Methods]  
{High statistics angular distributions and total cross sections for the photoproduction 
of $\pi^0$ mesons off the deuteron with coincident detection of recoil nucleons have been 
measured for the first time. The experiment was performed at the tagged photon beam of the 
Mainz MAMI accelerator for photon energies between 0.45~GeV and 1.4~GeV, using an almost 
$4\pi$ electromagnetic calorimeter composed of the Crystal Ball and TAPS detectors. 
A complete kinematic reconstruction of the final state removed the effects of Fermi motion.} 
\item[Results]
{Significant effects from final state interactions (FSI) were observed for participant
protons in comparison to free proton targets (between 30\% and almost 40\%). The data in 
coincidence with recoil neutrons were corrected for such effects under the assumption that 
they are identical for participant protons and neutrons.
Reaction model predictions and PWA for $\gamma n\rightarrow n\pi^{0}$, based on fits to data 
for the other isospin channels, disagreed between themselves and no model provided a good 
description of the new data.} 
\item[Conclusions]
{The results demonstrate clearly the importance of a mesurement of the fully neutral final state
for the isospin decmposition of the cross section. Model refits, for example from the Bonn-Gatchina 
analysis, show that the new and the previous data for the other three isospin channels can be 
simultaneously described when the contributions of several partial waves are modified. The results 
are also relevant for the suppression of the higher resonance bumps in total photoabsorption on nuclei,
which are not well understood. }
\end{description}
\end{abstract}

\maketitle

\section{Introduction}

\begin{figure*}[thb]
\centerline{\resizebox{1.0\textwidth}{!}{%
  \includegraphics{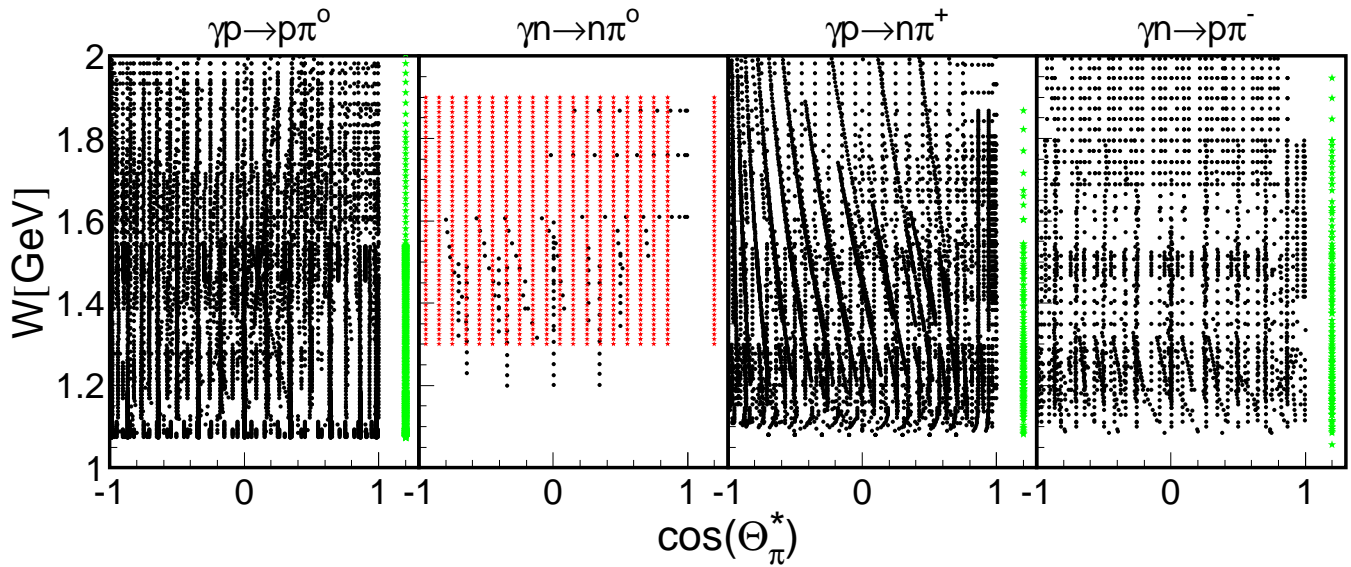}
}}
\caption{Data coverage for angular distributions and total cross sections
(green stars at cos($\theta_{\pi}^{\star}$) = 1.1) for the photoproduction of pions off 
the nucleon as a function of invariant mass $W$ and of pion momentum polar angle 
$\theta_{\pi}^{\star}$. Black circles: previous data, red stars: $n\pi^0$ final state results 
from this work. 
}
\label{fig:pion_base}       
\end{figure*}

The photoproduction of mesons is a prime tool for the study of the excitation spectrum 
of the nucleon, which is a major testing ground for the properties of the strong 
interaction in the non-perturbative regime. The pion is the lightest meson and has 
a strong coupling to many nucleon excited states. Although recent years have provided new 
photoproduction data for many different final states, pion scattering and photoproduction 
of pions are still central to most analyses which aim to identify and characterize the 
excited states of nucleons. Many theoretical frameworks are employed to extract this information. 
They include the SAID multipole analysis 
\cite{SAID,SAID_new}, the MAID unitary isobar model \cite{MAID,MAID_new}, 
the Dubna-Mainz-Taipei (DMT) dynamical model \cite{DMT}, the Bonn-Gatchina (BnGa) 
coupled channel analysis \cite{Anisovich_10}, the effective Lagrangian models of the Giessen 
group \cite{Feuster_97,Feuster_99} and the Madrid group \cite{Fernandez_06}, the 
J\"ulich-Bonn dynamical coupled channel analysis \cite{Roenchen_13}, the KSU model 
\cite{Shrestha_12}, and  the analysis of the recent CLAS data for the electroproduction 
of pions \cite{Aznauryan_09}. 

The database for pion photoproduction off the free proton is large and rapidly
growing, in particular for the $\gamma p\rightarrow p\pi^0$ reaction
\cite{Bartholomy_05,Bartalini_05,vanPee_07,Dugger_07,Elsner_09,Sparks_10,Crede_11,Thiel_12,Gottschall_14,Sikora_14,Schumann_15,Adlarson_15,Hartmann_15,Gardener_16,Annand_16,Thiel_17}
(references to data sets published before 2005 can be found in \cite{vanPee_07}),     
including results from the measurements of single and double polarization observables 
with CLAS at JLab, Crystal Barrel/TAPS at ELSA, Crystal Ball/TAPS at MAMI, and GRAAL at ESRF. 
However, a complete partial wave analysis necessitates the isospin decomposition of the 
electromagnetic excitations \cite{Krusche_03}. This requires the measurement of at least 
one pion production reaction off the neutron. The database for meson production reactions 
off the neutron, in particular for neutral pions, is significantly sparser than the proton data.
Historically, the difference arose due to the complications involved in measurements with 
quasi-free neutrons. However, many efforts are currently under way to improve this situation 
\cite{Krusche_11}.

The database for angular distributions of single pion production reactions off the nucleon which was
available when the present results were published as a letter \cite{Dieterle_14}
is summarized in Fig.~\ref{fig:pion_base}. In the meantime further data for the 
$\gamma n\rightarrow p\pi^-$ reaction have been published from the CLAS experiment 
\cite{Mattione_17,Ho_17}. The figure shows the kinematic ranges covered by the
previous data, binned in invariant mass $W$ and center of momentum (cm) angle $\theta_{\pi}^{\star}$ 
(plotted is cos$(\theta_{\pi}^{\star})$). Also shown are the present data points 
for the $\gamma n\rightarrow n\pi^0$ reaction, which had previously only been minimally investigated.  
Data for polarization observables for the $n\pi^0$ final state were also very sparse until recently. 
The beam asymmetry $\Sigma$ has been measured by the GRAAL collaboration \cite{DiSalvo_09} 
and first results for the double polarization observable $E$ measured with longitudinally polarized 
target and circularly polarized beam, were reported by the Crystal Ball/TAPS collaboration 
\cite{Dieterle_17} very recently. In the range of the $\Delta$ resonance, results for the helicity 
dependence of single pion production were also reported from the GDH experiment at MAMI \cite{Ahrens_10}, 
but mainly for charged pions and at photon energies lower than those in the present experiment.

The situation is better for $\gamma n\rightarrow p\pi^-$ since this final state can 
be detected with magnetic spectrometers. One might argue that the lack of data 
for the $n\pi^0$ final state is not a severe problem, since in principle the measurement 
of the other three isospin channels (see below) is enough to fix the three independent 
isospin amplitudes $A^{IS}$, $A^{IV}$, and $A^{V3}$ \cite{Krusche_03}. However, the predictions
of different reaction models and PWA for $\gamma n\rightarrow n\pi^0$ based on the results of the other
isospin channels differed widely \cite{Dieterle_14}. The main problem is that for the isospin
channels with charged pions, contributions from non-resonant backgrounds are much more 
important \cite{Krusche_03}. In the absence of complete data sets with a sufficient database 
of polarization observables \cite{Chiang_97}, significant model dependencies can exist. 

\begin{figure}[!htbp]
\centerline{ 
\resizebox{0.5\textwidth}{!}{%
  \includegraphics{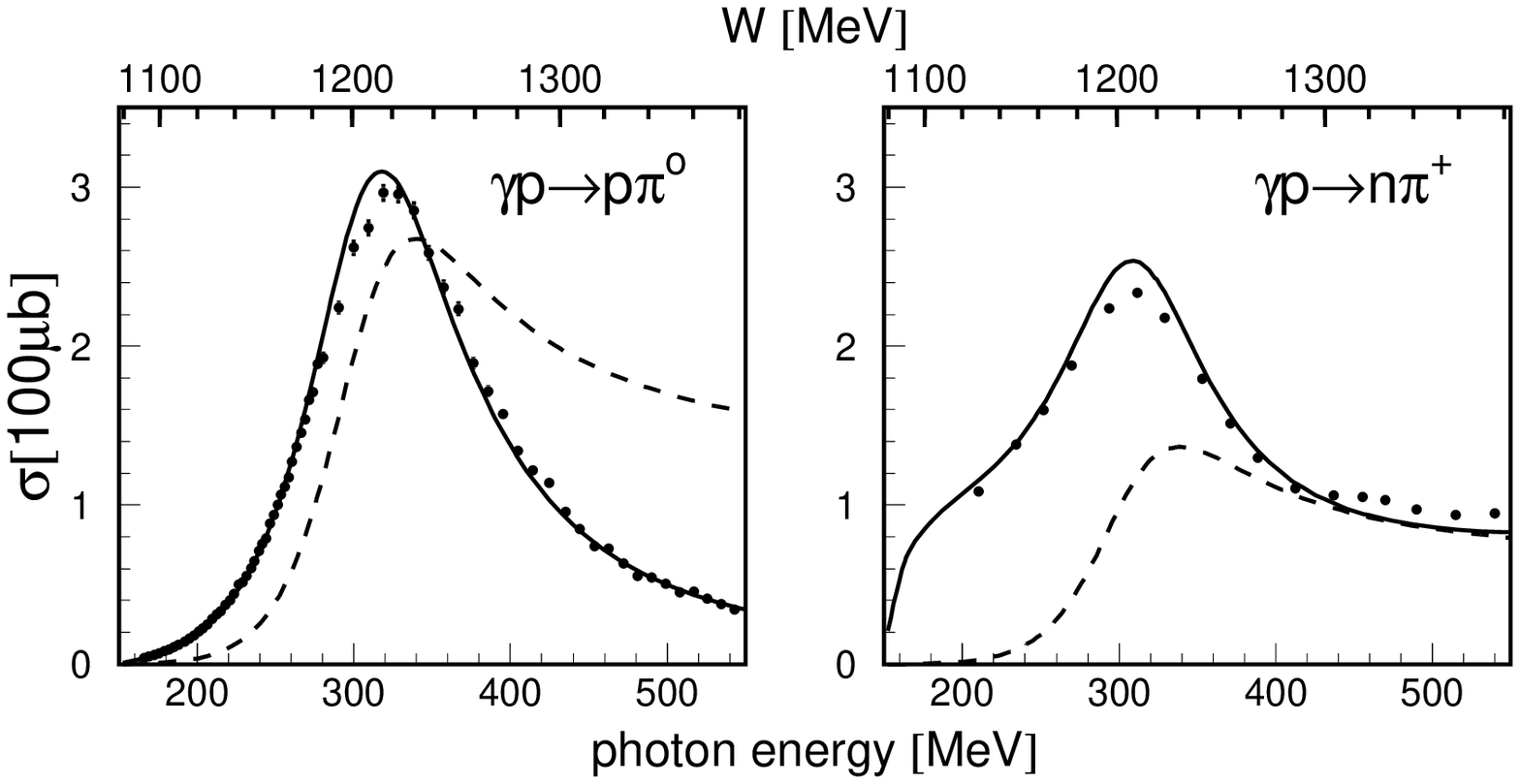} 
}}
\centerline{ 
\resizebox{0.5\textwidth}{!}{%
  \includegraphics{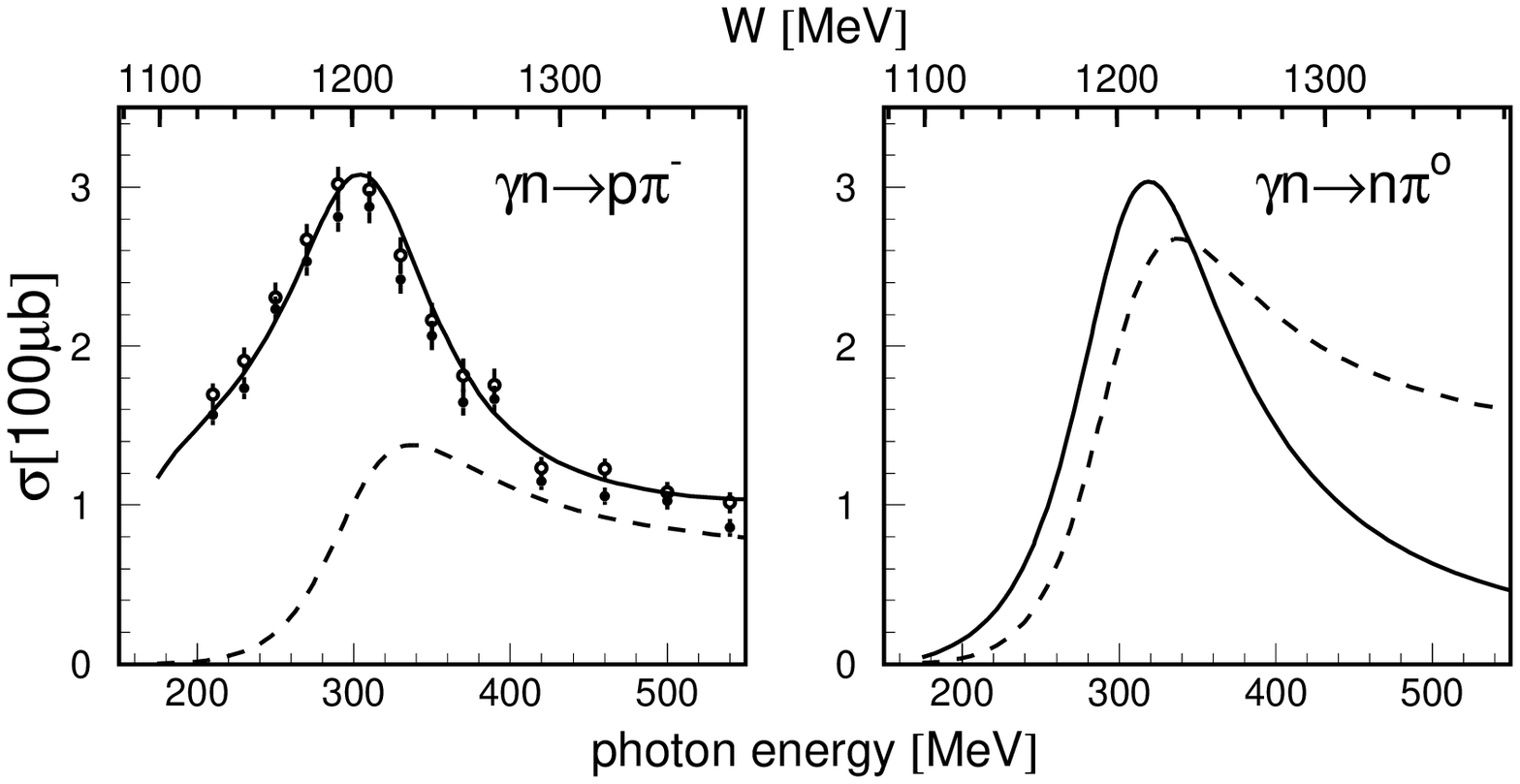}
}
}
\caption{Pion production in the $\Delta$-resonance region. 
Measured cross sections: $p\pi^0$ final state \cite{Fuchs_96,Krusche_99}, 
$n\pi^+$ final state \cite{Buechler_94},
$p\pi^-$ final state \cite{Benz_73}. Curves: MAID-model \cite{MAID}, solid: full model,
dashed: only P$_{33}$(1232) resonance.
}
\label{fig:delta}       
\end{figure}

The photoproduction of neutral pions has the advantage that background contributions, e.g. 
from Kroll-Rudermann or pion-pole terms, are suppressed because the incident photon 
cannot couple to the pion via its charge. A simple example is pion photoproduction 
in the $\Delta$-resonance region summarized in Fig~\ref{fig:delta}.
It follows immediately from the isospin decomposition that for pure excitation of the
$P_{33}$ resonance, without background contributions, the cross sections for the four
isospin channels are related by
\begin{eqnarray}
 \sigma(\gamma p\rightarrow p\pi^0) & =  
 \sigma(\gamma n\rightarrow n\pi^0) & = \nonumber\\
2\sigma(\gamma p\rightarrow n\pi^+) & =  
2\sigma(\gamma n\rightarrow p\pi^-) &, 
\end{eqnarray} 
which is obviously not the case for the experimental results. The reason is the large background
contribution to the reactions with charged pions in the final state. The MAID-model
results for the $P_{33}$ (dashed lines in the figure) respect this relation. However,  
roughly 50\% of the cross section for the charged channels at the $\Delta$
peak position are related to background contributions, which are even different for
the positively and negatively charged pions. Therefore, experimental data for the $n\pi^0$ 
channel are necessary for better control of the separation of resonance and background 
contributions in the reaction models.

\begin{figure}[thbp]
\centerline{ 
\resizebox{0.225\textwidth}{!}{%
  \includegraphics{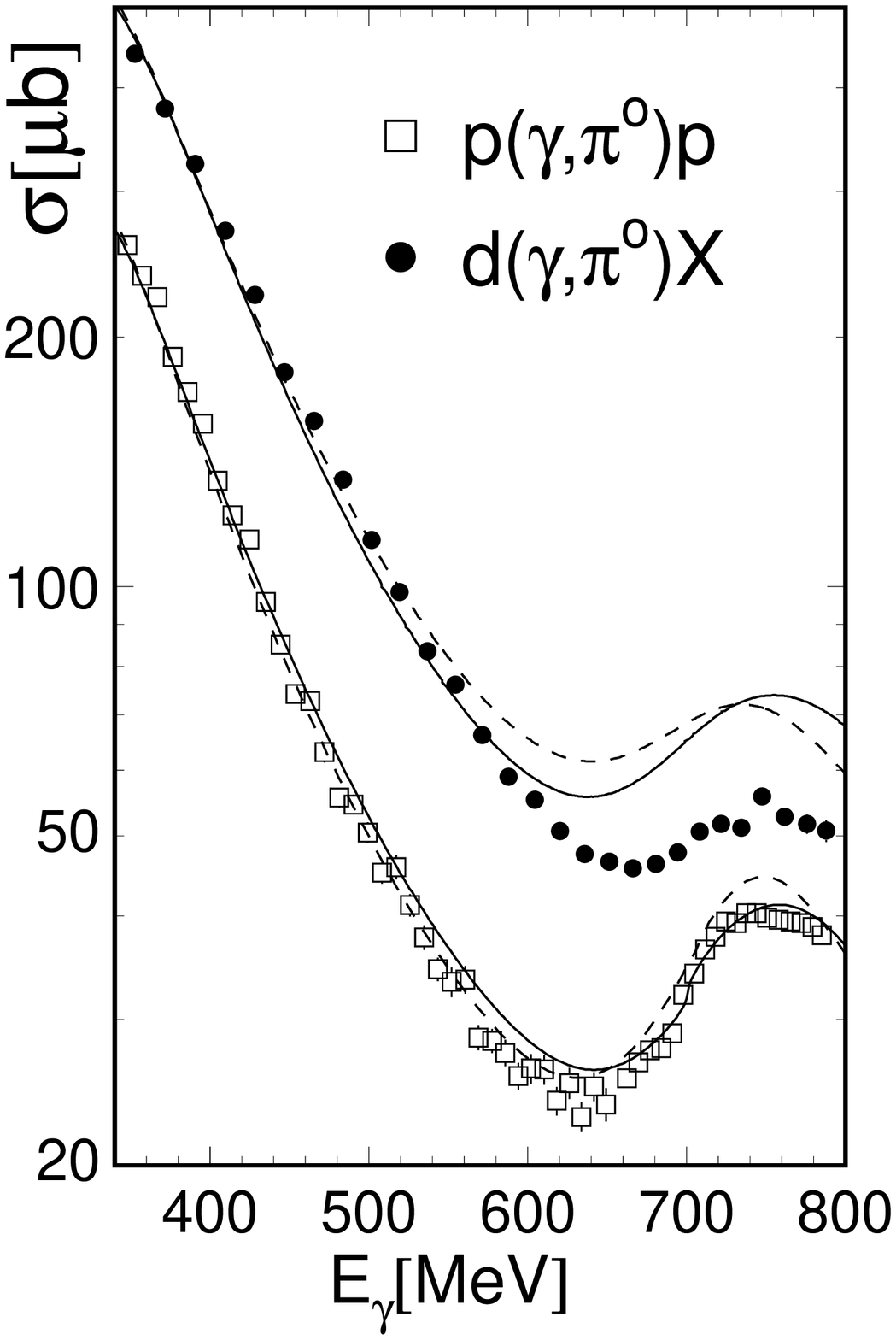} 
}
\resizebox{0.27\textwidth}{!}{%
  \includegraphics{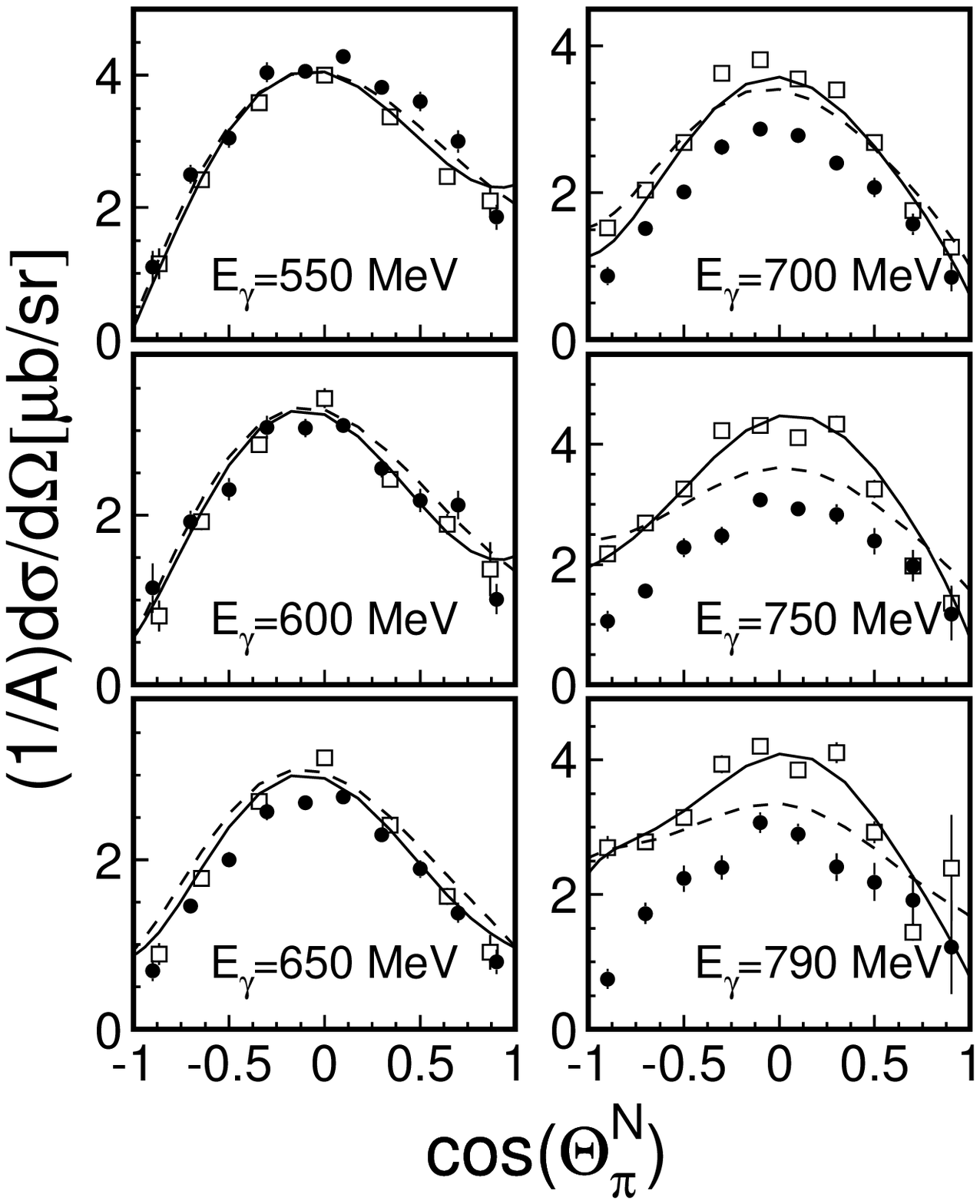}
}}
\caption{Single $\pi^0$ photoproduction off the free proton and the deuteron in
the second resonance region (note that $d(\gamma,\pi^0)X$ includes the $np\pi^0$
and $d\pi^0$ final states) \cite{Krusche_99}. Left hand side: total cross sections. 
Curves: results from the SAID analysis \cite{SAID} (solid), and MAID-model 
\cite{MAID} (dashed). For the deuteron from both models, the sum of
proton and neutron cross section folded with nuclear Fermi motion is plotted. 
Right hand side: angular distributions, solid curves: SAID proton, dashed curves:
Fermi smeared average of SAID proton and neutron.
}
\label{fig:secres}       
\end{figure} 

Measurements off quasi-free neutrons are complicated by nuclear Fermi motion and possible
nucleon-nucleon and nucleon-meson final state interaction (FSI) effects. The effects
from Fermi motion can be reliably removed (within experimental resolution) with a 
kinematic reconstruction of the final state invariant mass \cite{Krusche_11}. Thus, they are 
not problematic unless narrow structures in the cross section must be resolved. 
The importance of FSI effects can vary considerably for different final states. 
This can be tested with a comparison of the cross section data for free and quasi-free 
protons. Results for quasi-free photoproduction of $\eta$ and $\eta '$ mesons off the 
deuteron \cite{Jaegle_11,Jaegle_11a} show no significant FSI influence at the current 
level of the statistical precision of the experimental data.  However, results for the quasi-free 
$\gamma n\rightarrow p\pi^-$ reaction \cite{Mattione_17,Tarasov_11,Chen_12,Briscoe_12} found 
significant FSI effects, in particular for forward-meson angles. This is the kinematic regime where 
nucleon-nucleon FSI becomes important due to the small relative momentum between the `participant' 
and `spectator' nucleon. Also this complication makes it desirable to study both pion reaction 
channels off the quasi-free neutron, which will allow better approximations of such systematic 
effects.

In the case of $\pi^0$ photoproduction off the deuteron, the coherent process 
$\gamma d\rightarrow d\pi^0$ will contribute in addition to the breakup reaction
$\gamma d\rightarrow np\pi^0$. This contribution is large in the $\Delta$-resonance region,  
in particular for pion forward angles, and it removes strength from the quasi-free reactions
\cite{Krusche_99}.
The net effect is that the sum of the elementary cross sections for free protons and free 
neutrons - after folding with Fermi motion - is better approximated by the inclusive cross 
section for $\gamma d\rightarrow X\pi^0$ than by the sum of the exclusive quasi-free cross 
sections for $\gamma d\rightarrow p\pi^0(n)$ and $\gamma d\rightarrow n\pi^0(p)$. In the 
$\Delta$-resonance region, such effects have been studied in detail with models taking 
into account FSI and with experimental data comparing free and quasi-free production off protons
\cite{Darwish_03,Kossert_04}. The coherent contribution diminishes at higher 
incident photon energies due to the deuteron form factor.

Prior to this experiment, to our knowledge, no data for the exclusive quasi-free reactions 
$\gamma d\rightarrow (n)p\pi^0$, $\gamma d\rightarrow n(p)\pi^0$ (in parentheses: spectator nucleon)
existed. There are, however, some results for the inclusive reaction $\gamma d\rightarrow X\pi^0$
\cite{Krusche_99} up to the second resonance region (see Fig.~\ref{fig:secres}). The second 
resonance peak is less prominent in these data than for free protons. The Fermi smeared sum of the 
results of the SAID \cite{SAID} and MAID \cite{MAID} models for the elementary reactions on protons 
and neutrons agreed with the measured cross section in the tail of the $\Delta$-resonance, but 
overestimated the second resonance peak. It was unclear whether this indicated a problem of the 
models for the neutron cross section, large FSI effects, or both. Only an exclusive measurement 
with coincident recoil nucleons could clarify this.

The present work summarizes the results from a measurement of single $\pi^0$ photoproduction 
off the deuteron with detection of the pion-decay photons and the recoil nucleons for incident 
photon energies from $\approx$ 450~MeV to 1400~MeV. The paper is organized in the following way:
A short description of the experimental setup is given in Section \ref{sec:setup}. The different 
steps of the analysis are discussed in Section \ref{sec:ana}. In Section \ref{sec:results}, 
we first discuss the results for the quasi-free processes as a function of incident photon energy  
(i.e. cross sections folded with nuclear Fermi motion) and subsequently the results as function of 
final state invariant mass, which can be compared to previous experimental data for the proton target 
and to model predictions for the free cross sections for protons and neutrons. Some of the results 
have already been published in a letter \cite{Dieterle_14}. 
This paper gives more details about the analysis and presents also results which could not be 
included in the letter (e.g. the experimental data without corrections for Fermi motion).  

\section{Experimental setup}
\label{sec:setup}

The experiment was performed at the electron accelerator facility MAMI in Mainz 
\cite{Herminghaus_83,Walcher_90,Kaiser_08} using a quasi-monochromatic photon beam with 
energies between $\approx$0.45 GeV and $\approx$1.4 GeV from the Glasgow tagged photon 
spectrometer \cite{Anthony_91,Hall_96,McGeorge_08}. In total, three beam times with a liquid 
deuterium target were taken (see \cite{Oberle_14,Werthmueller_14,Dieterle_15,Kaeser_16} 
for details). One of them, optimized for multiple meson production, used a trigger with 
hit multiplicity three and was not analyzed for the present results. The two beam times 
analyzed here used primary electron beams with energies of 1.508~GeV and 1.557~GeV which 
produced bremsstrahlung in a copper radiator of 10 $\mu$m thickness. The typical energy 
resolution of the photon beam was defined by the 4 MeV bin width of the tagger focal plane 
detectors. The electron beam was longitudinally polarized so that the photon beam was 
circularly polarized. This was, however, irrelevant for the present results since the target 
was unpolarized and single-meson production from an unpolarized target shows no asymmetries
for a circularly polarized beam due to parity conservation. The polarization degree of freedom 
was used in the analysis of the production of meson pairs ($\pi^0\pi^{0,\pm}$, $\pi^{0,\pm}\eta$), 
which were measured simultaneously \cite{Oberle_13,Oberle_14,Kaeser_16}. 

\begin{figure}[!thbp]
\centerline{\resizebox{0.50\textwidth}{!}{%
  \includegraphics{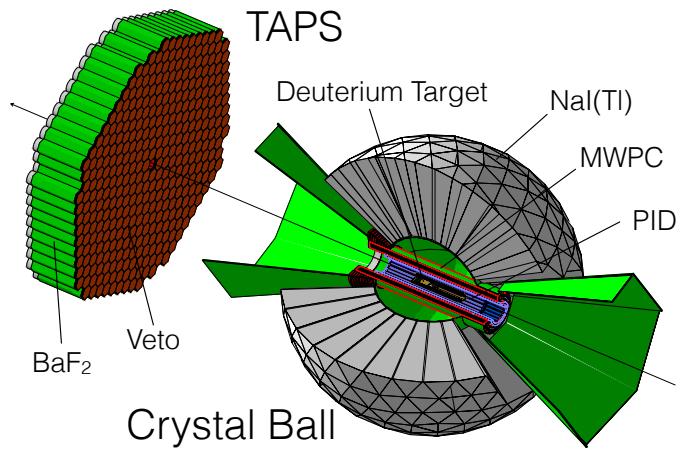}
}}
\caption{Setup of the electromagnetic calorimeter combining the Crystal Ball and TAPS 
(left-hand side) detectors. Only three quarters of the Crystal Ball are shown. 
Detectors for charged particle identification were mounted in the Crystal Ball 
(PID and MWPC) and in front of the TAPS forward wall (TAPS Veto-detector, CPV). 
The beam enters from the bottom right corner of the figure.
}
\label{fig:calo}       
\end{figure}

The target material was liquid deuterium contained in Kapton cylinders of $\approx$ 4~cm 
diameter and 4.72~cm or 3.02~cm length corresponding to surface densities of 
0.231 nuclei/barn or 0.147 nuclei/barn, respectively. The beam spot size on the target 
($\approx 1.3$ cm diameter) was defined by a collimator (4 mm diameter) placed downstream 
from the radiator foil. The photon flux, needed for the absolute normalization of 
the cross sections, was derived from the number of deflected electrons and the 
fraction of correlated photons that pass the collimator and reach the target (tagging efficiency). 
The flux of scattered electrons was counted by live-time gated scalers. The tagging efficiency 
was determined with special experimental runs. A total absorbing lead-glass counter was moved 
into the photon beam at reduced intensity of the primary electron beam. In addition to
these periodical absolute measurements, the photon beam intensity was monitored in arbitrary 
units during normal data taking with an ionization chamber at the end of the photon-beam
line.

Photons and recoil nucleons were detected using an almost $4\pi$ electromagnetic 
calorimeter, supplemented with detectors for charged particle identification 
(see Fig.~\ref{fig:calo}). More details of the calorimeter (in a slightly different 
configuration) are given in \cite{Schumann_10,Zehr_12}. The setup combined the Crystal 
Ball (CB) detector \cite{Starostin_01} with a hexagonal forward wall constructed from 
384 BaF$_2$ modules from the TAPS array \cite{Novotny_91,Gabler_94}. Between the two beam
times, TAPS was modified by replacing the two inner-most rings close to the beam pipe 
by trapezoidally shaped PbWO$_4$ crystals (four crystals for each BaF$_2$ module) to increase 
rate capability. However, these new modules were not yet operational and were not used in 
the analysis. The Crystal Ball is made of 672 NaI detectors, arranged in two half spheres, which 
together cover the full azimuthal range for polar angles from 20$^{\circ}$ to 160$^{\circ}$, 
corresponding to 93\% of the full solid angle. The TAPS forward wall was placed 1.468~m downstream 
from the target and covered polar angles between $\approx$5$^{\circ}$ and 21$^{\circ}$. 
All TAPS modules were equipped with individual plastic scintillators (Charged Particle Veto, CPV) 
in front of the crystals for charged particle identification. The target cell with the liquid 
deuterium was mounted from the upstream side with its cryo-support structures in the center of 
the CB. It was surrounded by a detector for charged particle identification (PID) \cite{Watts_05} 
and multiwire-proportional chambers (MWPC) which were fitted into the beam tunnel of the CB. 
The MWPC for charged particle tracking were not used in the present analysis.
The PID consisted of 24 plastic scintillators, which surrounded the target and provided full 
azimuthal coverage. Each scintillator covered $15^{\circ}$ of azimuthal angle and the same 
range in polar angle as the CB, i.e. from $20^{\circ}$ to $160^{\circ}$. The PID did not
provide polar angle information.

For trigger purposes, the CB and TAPS were subdivided into logical sectors. The CB was split
into 45 rectangular areas (after projecting its geometry on a plane) and TAPS into 6$\times$64 
modules in a pizza-slice geometry.
The trigger condition used for the present analysis was a multiplicity of two logical
sectors with the signal of at least one detector module above a threshold of about 30 MeV (CB) 
or 35 MeV (TAPS) and the analog energy-sum signal from the CB above 300 MeV. This condition 
was not optimized for the measurement of single $\pi^0$ production, but for the simultaneous 
measurement of $\eta$- and multiple meson production reactions. Events with both photons 
going into TAPS were not accepted. In the analysis, only events were used for which these 
conditions were fulfilled already by the $\pi^{0}$-decay photons. Events where the trigger was only 
activated due to the additional energy deposition of the recoil nucleon were discarded in order 
to avoid systematic uncertainties (the energy response of the detector was calibrated for 
photon showers, not for recoil nucleons). For accepted events, the readout thresholds 
for the detector modules were set to 2~MeV for the CB crystals, to 3-4~MeV for the TAPS 
crystals, to 250~keV for the TAPS charged-particle scintillators, and to 350~keV for 
the elements of the PID. 

\section{Data analysis}
\label{sec:ana}
The data used for the present analysis were also used to investigate several other meson 
production reactions ($\eta$-mesons \cite{Werthmueller_13,Werthmueller_14}, $\pi\pi$ pairs 
\cite{Oberle_13,Oberle_14,Dieterle_15}, and $\pi\eta$ pairs \cite{,Kaeser_15,Kaeser_16}). 
The reliability of the raw data, of the calibration procedures, and of the analysis 
strategies was tested in several independent ways and details have been given in the above 
mentioned publications. Therefore, only a summary of the main analysis steps and specific 
details for the analysis of the $\gamma N\rightarrow N\pi^0$ reactions with quasi-free 
nucleons are given here. 

The analysis was based on five main steps: (1) the calibration of all detector elements in use
(Crystal Ball, TAPS, PID, CPV, and tagging spectrometer) in view of energy and/or timing information,
(2) the identification of events from the $\gamma N\rightarrow N\pi^0$ reaction (particle
identification, invariant, and missing mass analyses etc.), (3) the absolute normalization of 
the cross sections (beam flux, target density, and Monte Carlo simulations of the detection 
efficiency), (4) the reconstruction of the total cm energy $W$ from the final-state kinematics 
for events in which the effects of Fermi motion were removed, and (5) the correction for FSI 
for the quasi-free neutron results.  

\subsection{Detector Calibration}

A detailed description of the detector performance and the calibration procedures was already given 
in \cite{Schumann_10,Zehr_12,Witthauer_13,Werthmueller_14,Kaeser_16}. Timing information was available 
for the plastic scintillators of the focal plane (FP) detector of the tagging spectrometer, the NaI
crystals of the CB, the BaF$_2$ modules of TAPS, the plastic scintillators of the PID detector, 
and the scintillators from the TAPS veto detector. The CB and the FP detector were equipped with
CATCH TDCs of a fixed conversion gain of 117~ps/channel. The gains of the TAPS modules were 
calibrated by inserting delay cables of precisely known lengths into the common stop signal.
The offsets (time-zero-position of the signals) were calibrated by iterative procedures comparing 
coincident signals within and between different detector components. The slow signals from the CB 
detector, analyzed with Leading Edge Discriminators (LED), required in addition an energy dependent 
time-walk correction, which greatly improved time resolution. In contrast, the fast signals from the 
TAPS detector analyzed with Constant Fraction Discriminators (CFD) needed no time-walk correction. 
Typical time resolutions (time spectra are e.g. shown in \cite{Witthauer_13,Werthmueller_14})
with this setup are listed in Tab.~\ref{tab:time_resol}.

\begin{table}[hhh]
\begin{center}
  \caption[Time resolutions]{
    \label{tab:time_resol}
     Typical time resolutions (FWHM) for coincidences between different detector components.
}
\vspace*{0.3cm}
\begin{tabular}{|c|c|}
\hline
Detector coincidence & typical resolution [ns]\\
\hline\hline
  $TAPS - TAPS$  & 0.45 - 0.55 \\
  $TAPS - CB$     & 1.3 - 1.0 \\
  $CB - CB$ & 2.0 - 3.0\\
  $TAPS - Tagger$ & 0.8 - 1.0\\ 
  $CB - Tagger$ & 1.4 - 1.6\\    
\hline
\end{tabular}
\end{center}
\end{table}

Most important were the CB-Tagger and TAPS-Tagger time resolution because the size of the 
background from random tagger - production-detector coincidences depends on it. The random 
background was removed in the usual way by a side-band subtraction in the time spectra 
(see e.g. \cite{Witthauer_13,Werthmueller_14}). Furthermore, the timing information from the 
TAPS detector was important for a time-of-flight (ToF) versus energy analysis for the separation of 
different particle types in the TAPS forward detector. The CB-CB timing information and 
the timing informations from the PID and TAPS CPV were only used to assure that hits in these 
detectors corresponded to the same event. However, the background from event overlap was anyway 
negligible, so that time resolution was not an important issue in this case.
Energy information was available from the modules of the CB and TAPS calorimeters and the 
PID and TAPS CPV devices. For the photon-tagger, energy information came not from the response 
of the FP scintillators, but from their geometric position in the focal plane calibrated by 
special measurements \cite{McGeorge_08} with direct deflection of electron beams of precisely known
energies into the focal plane. 

The primary pre-data-taking calibration of TAPS was done with cosmic muons, which (as minimum 
ionizing particles) deposit on average approximately 37.7~MeV per crystal because, in contrast 
to the CB, all crystals have the same geometry and are horizontally oriented in the same way. 
A rough energy calibration of the CB was done before data taking with an $^{241}$Am/$^9$Be source 
(photons of 4.438 MeV and a continuous neutron spectrum up to about 10 MeV) placed at 
the target position. 

The final calorimeter calibration started with the CB. In an iterative procedure, 
the invariant mass of photon pairs identified as decay products of $\pi^0$ mesons was first 
used for a linear calibration. This was subsequently improved by a quadratic term derived 
from the invariant mass of photon pairs from $\eta$-meson decays. 
The energy response of the TAPS detector was calibrated in the same way. However, since 
two-photon hits in TAPS are rare for $\pi^0$ decays and almost impossible for $\eta$ decays, 
events with one photon in CB and one photon in TAPS had to be used. Therefore, the TAPS 
calibration depends on the previous CB calibration. Furthermore, the scintillation light 
from BaF$_2$ crystals has two different components with different wavelengths, decay times,
and relative intensities depending on the type of the detected particle 
\cite{Novotny_91,Gabler_94}. This feature is routinely exploited by a pulse-shape analysis (PSA)
used for particle identification by integrating the signals over a short and a long gate period. 
Therefore, two independent energy signals had to be calibrated for TAPS. As usual, the calibration 
was done in a way that the calibrated short-gate and long-gate energy signals were identical for 
photons.

The energy response of the PID detector was calibrated by a comparison of
the $E-\Delta E$ spectra measured for clearly identified protons to the results from Monte Carlo 
simulations. The energy signals of the CPV were not further used in the analysis, their 
calibration was only relevant for the determination of the correct veto thresholds. This was 
also done by comparison to Monte Carlo simulations.

\subsection{Particle Identification}  

\begin{figure*}[!thbp]
\centerline{\resizebox{1.0\textwidth}{!}{%
  \includegraphics{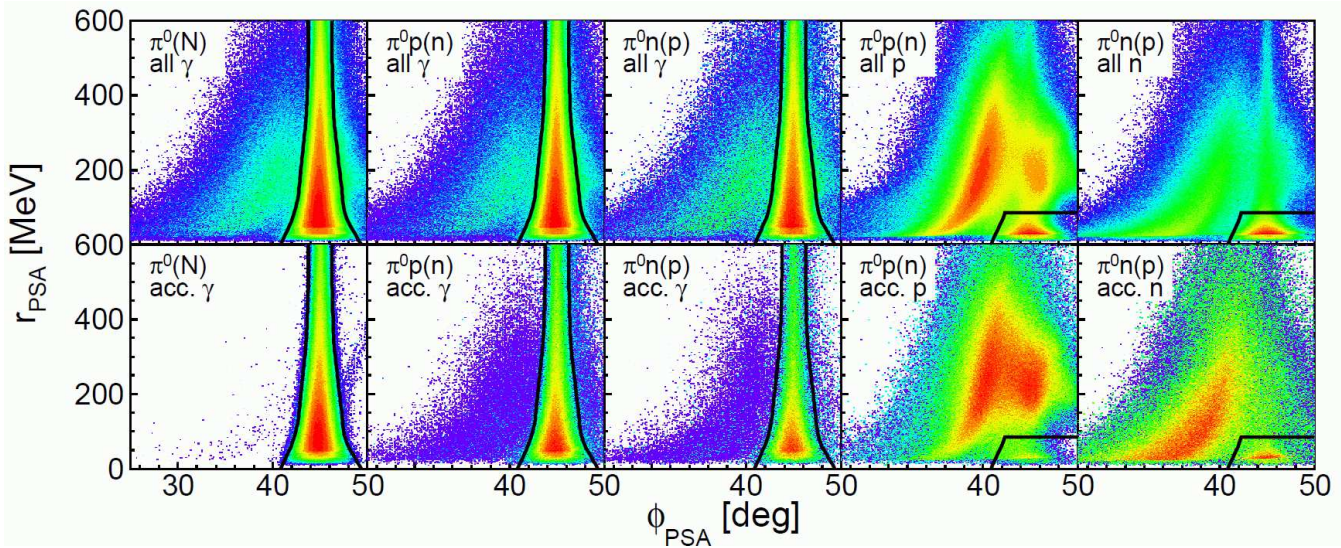}
}}
\caption{PSA spectra for hits in TAPS. 
Top row: raw spectra selected with information from CPV detector and $\chi^2$ analysis 
(where applicable). From left to right: photon candidates for inclusive analysis 
(no condition for recoil nucleons), photons with coincident proton candidates,
photons with coincident neutron candidates, 
candidates for recoil protons, candidates for recoil neutrons.
Bottom row: same after application of all kinematic cuts. 
The black lines show the cuts applied to the spectra.} 
\label{fig:raw_psa}       
\end{figure*}

All results shown in this section were integrated over the full tagged and analyzed energy 
range of $E_{\gamma}$ from 0.45 - 1.4 GeV.
In the first step of the analysis, all modules of the main detectors CB and TAPS that detected 
a signal were grouped into connected clusters corresponding to hits from photons or massive
particles in the calorimeter. The position, time, and energy information of the clusters 
were then derived by summing up or averaging over the signals from the activated 
crystals \cite{Gabler_94,Schumann_10}. The position (i.e. the polar angle information) from 
clusters in the TAPS forward wall had to be corrected for the geometrical effect arising 
because the crystals arranged in a horizontal position were not pointing directly towards 
the target. This is a straight forward analytical correction, which only requires knowledge 
about the (energy dependent) average depths of the energy deposition in the detector. 
Subsequently, the clusters were assigned to the two types `neutral' or `charged' depending, 
for the CB, on the response of the PID and, for TAPS, on the response of the CPV. 
For the CB, hits were assigned as `charged' when the PID registered a coincident hit between 
the central CB-cluster module and the PID-scintillator bar within an azimuthal angle of 
$15^{\circ}$. For TAPS, a hit was assigned as `charged' when the CPV element in front of 
the central cluster module or a CPV neighbor module of the central cluster module responded. 
Due to the horizontal arrangement of the TAPS modules, especially at larger polar angles, 
a charged particle may not pass the central CPV, but the neighboring module at a different 
polar angle.     
   
Three different types of events were analyzed for the present work. Events with exactly two
neutral and one charged hit were accepted as candidates for the exclusive 
$\gamma d\rightarrow (n)p\pi^0$ reaction ($\sigma_p$, $\pi^0$ and participant proton). 
Events with exactly three neutral hits were analyzed for the exclusive $\gamma d\rightarrow (p)n\pi^0$ 
reaction ($\sigma_n$, $\pi^0$ and participant neutron). `Participant' proton (or neutron) were assigned
as the nucleon detected in coincidence with the pion. In rare cases, due to Fermi momenta in the  
tail of the bound-nucleon momentum distribution, also detection of the `spectator' nucleon was possible.
This was included into the MC simulations of detection efficiency; only second order effects from FSI 
modifying the tail of the distributions could not be accounted for. In addition, the inclusive reaction 
$\gamma d\rightarrow X\pi^0$ ($\sigma_{\rm incl}$) was analyzed, where $X$ corresponded to
a charged, a neutral, or no third hit in the calorimeter. This sample included events for 
which the recoil nucleon was not detected (if it was detected, it was ignored in the analysis)
and also events from the $\gamma d\rightarrow d\pi^0$ reaction. This inclusive analysis was independent 
of recoil nucleon detection efficiencies.

For all events with three neutral hits, the most probable assignment of them to the two 
$\pi^0$-decay photons and a neutron candidate was determined by a $\chi^2$ test for which 
the invariant masses of all pairs of neutral hits were compared to the nominal mass $m_{\pi^0}$
of the $\pi^0$ meson
\begin{equation}
\chi^2(\gamma_i ,\gamma_j) = \left(\frac{m_{\gamma_i ,\gamma_j} - m_{\pi^0}}{\Delta m_{\gamma_i ,\gamma_j}}\right)^2\;,
\end{equation}
where $m_{\gamma_i ,\gamma_j}$ is the invariant mass of neutral hits $i$ and $j$, $1\leq i,j\leq 3$,
$i\neq j$ and $\Delta m_{\gamma_i ,\gamma_j}$ is their uncertainty computed from the experimental 
energy and angular resolution (determined with MC simulations). Only the best combination was kept 
for further analysis. This applied to the events analyzed for $\sigma_n$ and the subset of events 
for $\sigma_{\rm incl}$ with three neutral hits. 

Further methods of particle-type identification were available for the TAPS forward wall, where 
they were important to distinguish recoil nucleons (which were mostly detected in the angular 
range covered by TAPS) from photon showers. A very efficient particle identification in TAPS was 
based on the PSA of the signals from the BaF$_2$ crystals. The scintillation light from BaF$_2$ 
crystals is composed of two components with different wave lengths and different decay constants, 
$\tau=0.9$~ns for the `fast' component and $\tau=650$~ns for the `slow' component. The relative 
intensity of the two components is different for electromagnetic showers induced by photons 
(or electrons) and stopped massive particles such as recoil protons and neutrons. Therefore, 
the signals were integrated over two ranges (short gate: 40~ns, long gate: 2~$\mu$s). The first 
integral added the fast component and a small fraction of the slow component and the second 
contained the total signal. Both signals were calibrated for photon energies, so that the 
short ($E_s$) and long gate ($E_l$) signals for photon hits were equal. For massive particles, 
$E_s$ is then smaller than $E_l$. Instead of comparing $E_s$ and $E_l$, it is more 
convenient to use a transformation to the PSA radius $r_{\mathrm{PSA}}$ and the PSA angle 
$\phi_{\mathrm{PSA}}$ defined by:  
\begin{equation}
r_{\mathrm{PSA}} = \sqrt{E_s^2 + E_l^2} \quad \mathrm{and} \quad \phi_{\mathrm{PSA}} 
= \arctan (E_s / E_l).
\end{equation}
In this representation, photon hits appear at $\phi_{\mathrm{PSA}}\approx 45^{\circ}$ 
independent of $r_{\mathrm{PSA}}$ and recoil nucleons are located at smaller angles.
Figure~\ref{fig:raw_psa} summarizes typical PSA spectra. In the upper row, raw spectra
are shown, for which hits have only been characterized as photons, protons, or neutrons 
by the response of the CPV and the $\chi^2$ analysis of events with three neutral hits. 
The photon candidates are shown separately for reactions with no condition for recoil 
nucleons and for coincident protons and neutrons. The bottom row of the figure shows the 
same spectra after the application of the subsequent kinematic cuts (see 
Sec.~\ref{sub:reaction}). The photon sample was already quite clean for the raw data and 
application of the kinematic cuts removed most of the background. For the final analysis, 
an energy dependent $3\sigma$ cut, indicated in the figure, was applied to these spectra. 
For the recoil nucleons, some background from abundant electromagnetic processes survived 
all other cuts (visible at $\approx45^{\circ}$ and small $r_{\rm PSA}$) and was cut away 
in the PSA spectra. The spectrum for recoil neutrons was cleaned by the subsequent 
kinematic cuts, which removed events with three neutral hits for which the $\chi^2$ 
assignment to photon and neutron hits was incorrect. The spectrum for recoil protons 
showed also in the region of expected photon hits 
($\Phi_{PSA}\approx 45^{\circ}$, $r_{PSA}$ between 200 - 350 MeV) a significant 
structure. However, this is not background, but due to high energy protons which were not 
stopped in TAPS, but punched through the detector (protons can be stopped in TAPS only up 
to kinetic energies of $\approx$400~MeV). The difference in the shape of the BaF$_2$ 
signals for heavy charged particles compared to electromagnetic showers is due to the depletion 
of electronic bands in the  scintillator material close to the endpoint of the tracks of such 
particles. Therefore, punch-through protons not stopping in the scintillator produce signal 
shapes similar to photons. This effect is less pronounced for recoil neutrons, which, when 
not stopped by nuclear reactions, are usually not detected at all.   

\begin{figure}[htbp]
\centerline{\resizebox{0.5\textwidth}{!}{%
  \includegraphics{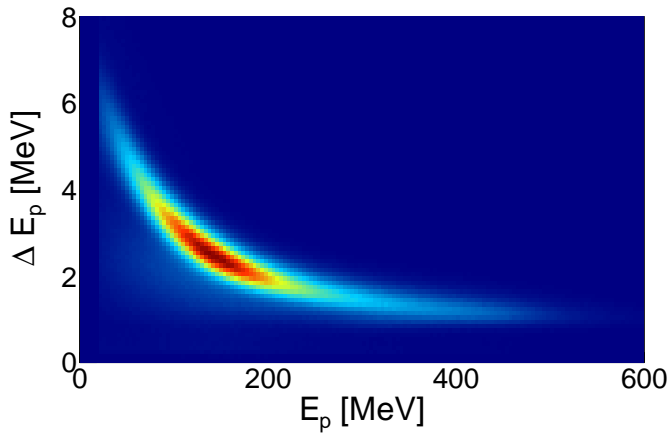}
}}
\caption{Proton identification by the CB - PID detector system. Shown is the energy loss 
$\Delta E_p$ in the PID versus the total deposited energy $E_p$ in the CB for hits identified
as protons, after all other analysis cuts. No background from electrons or charged pions is 
visible.
}
\label{fig:de_e}       
\end{figure}

Further particle identification methods were based on $E-\Delta E$ analyses comparing the 
energy loss of charged particles in the PID (CPV) detectors to the total deposited
energy in the CB (TAPS). 
The final result of the $E-\Delta E$ analysis for the CB-PID system is shown in 
Fig.~\ref{fig:de_e}. This spectrum shows a clean, background free signal for recoil protons.
Signatures for charged pions and deuterons were only visible in the raw spectra
(not shown here, see e.g. Ref. \cite{Oberle_14}) before application of the other cuts.
The resolution for the corresponding analysis using the CPV-TAPS system was less good
because, due to the readout with thin scintillating fibers, the light output from the CPV was low
so that the energy resolution was worse than for the PID. Typical spectra for the same data set 
but from an analysis of the $\eta\rightarrow 2\gamma$ and the 
$\eta\rightarrow 3\pi^0\rightarrow 6\gamma$ 
decays are shown in \cite{Werthmueller_14}. That analysis was not used here.

\begin{figure}[thbp]
\centerline{\resizebox{0.5\textwidth}{!}{%
  \includegraphics{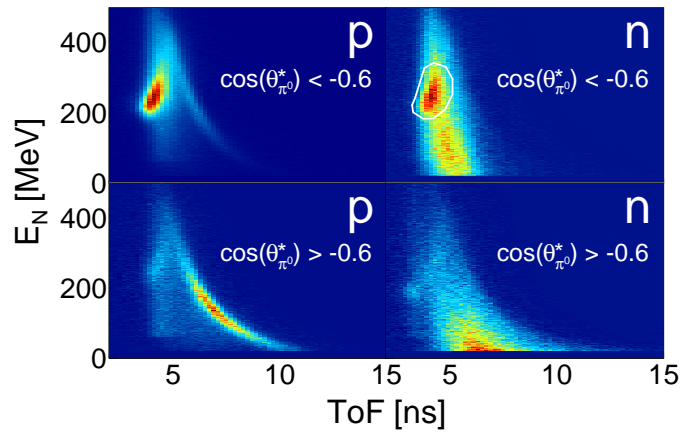}
}}
\caption{Nucleon identification with the TAPS detector showing the deposited 
energy of the nucleon $E_N$ versus its ToF (normalized to 1~m flight distance). 
Left column: proton, right column: neutron, 
top row: $\cos(\theta^{*}_{\pi^{0}})<-0.6$, bottom row: $\cos(\theta^{*}_{\pi^{0}})>-0.6$.
The white line in the upper right histogram indicates background events from misidentified
punch-through protons.}
\label{fig:tof}       
\end{figure} 

\begin{figure}[bhtp]
\centerline{\resizebox{0.5\textwidth}{!}{%
 \includegraphics{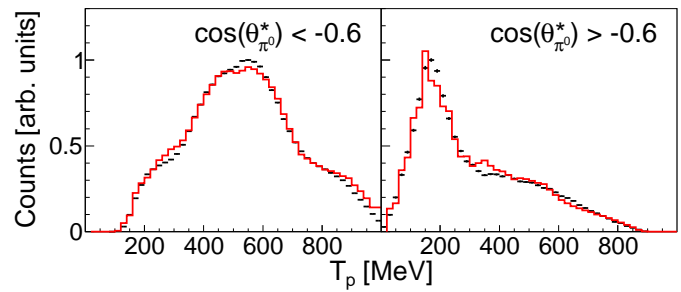}
 }}
 \caption{Kinetic energy distribution of the recoil proton for exclusive single $\pi^{0}$ 
 photoproduction off quasi-free protons for two different regions of 
 $\cos(\theta^{*}_{\pi^{0}})$. Black dots with error bars: Measured data, red line: MC signal.}
 \label{fig:kin}
 \end{figure}

\begin{figure*}[thbp]
\centerline{\resizebox{1.0\textwidth}{!}{%
  \includegraphics{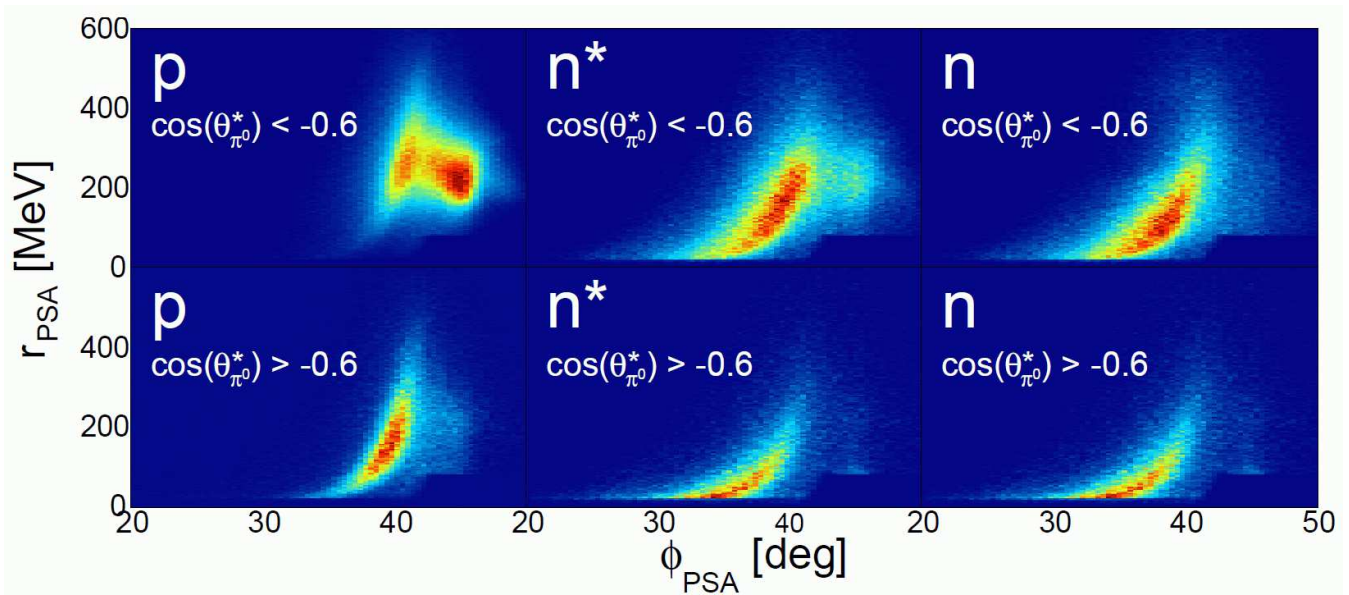}
}}
\caption{PSA analysis of hits in the TAPS detector for nucleon candidates for events
with forward and backward pion angles. Plotted is the PSA radius ($r_{\rm PSA}$ versus the
PSA angle ($\phi_{\rm PSA}$). 
Left column: proton, center column: neutron without ToF-versus-energy cut, 
right column: neutron with ToF-versus-energy cut. 
Top row: $\cos(\theta^{*}_{\pi^{0}})<-0.6$, bottom row: $\cos(\theta^{*}_{\pi^{0}})>-0.6$.}
\label{fig:psa}       
\end{figure*}

Due to the good time resolution of the TAPS detector and the 
relatively long flight path between the target and detector ($\approx$1.5~m), the comparison 
of the time-of-flight to the total deposited energy was also a powerful method to assign
hits in TAPS to different particle types. Spectra for proton and 
neutron candidates for two different angular ranges of the pions are shown in
Fig.~\ref{fig:tof}. Protons should appear in a relatively sharp band given by the 
relativistic velocity-energy relation. This was more or less the case for protons coincident
with pions going to forward angles, which correspond to low proton laboratory energies.
However, a small back-bending structure was visible already for this sample, corresponding
to punch-through protons which did not deposit their full energy in TAPS. This structure was
much more pronounced for pions at backward angles, for which a large number of protons were
high-energy, minimum-ionizing particles. No cuts were applied to the proton spectra.
Typical kinetic energy distributions (from kinematic reconstruction of the events) of the protons 
corresponding to the two different ranges of pion-cm angles are shown in Fig.~\ref{fig:kin}. 
Experimental results are compared to the output of the Monte Carlo simulations discussed in 
subsection~\ref{sub:MC}. 

With one exception discussed below, it was not necessary to apply cuts to the 
corresponding spectra. The background level in these spectra was already very low
after the neutral/charged selection with the PID and CPV, the TAPS PSA cuts, the $\chi^2$ 
analysis, and the kinematic cuts discussed in subsection \ref{sub:reaction}. 

Recoil neutrons can deposit any fraction of their kinetic energy in the detector and 
their signals are distributed over a large area in the ToF-versus-energy spectra.
The neutron spectrum coincident with pions at cos($\theta^{\star}_{\pi^0})>-0.6$ 
in Fig.~\ref{fig:tof} shows the expected behavior without any residual trace from the 
proton band, which would indicate misidentified protons. The neutron spectrum coincident
with pions at cos($\theta^{\star}_{\pi^0})<-0.6$ is less clean. It shows a significant
structure from high energy, minimum-ionizing protons which escaped detection from the CPV. 
The cut indicated by the white line in the figure was applied to remove this background.
This cut was also applied to the data from the MC simulations for the detection efficiency 
(see subsection~\ref{sub:MC}). 

After this cut, the PSA spectra for protons and neutrons were inspected again for
the two ranges of pion polar angles. The result is summarized in Fig.~\ref{fig:psa}.
The contribution of punch-through protons for backward pion angles is visible.
For smaller pion angles, some intensity at PSA angles $>45^{\circ}$ from punch-through
protons is also visible. The cut on ToF-versus-energy removed most background in this region
in the neutron spectra. The only cuts applied to these spectra were as indicated in
Fig.~\ref{fig:raw_psa} (i.e. in the extreme lower right corners of the spectra).

For the separation of photon and neutron hits in the CB, only the $\chi^2$ method could be 
used. Independent checks can be done with the analysis of the cluster multiplicity (i.e.
the average number of activated crystals per hit in the detector), which is smaller for 
neutrons than for photons. This has been tested with the same data set for the analysis 
of $\eta$-decays into two and six photons \cite{Werthmueller_14}. No indication for a
significant cross contamination was found, but the method does not allow a stringent
separation on an event-by-event basis, unless one accepts a large reduction of the
statistical quality of the data by only accepting multiplicity-one hits as neutrons.
No cuts were applied to cluster multiplicity in the present analysis.   

\subsection{Monte Carlo Simulations}
\label{sub:MC}

A reliable MC simulation of the response of the detector to the signal events is crucial for 
the absolute normalization of the experimental data. However, a comparison of signal and 
background events filtered through the detector response is also needed for the selection of 
the most efficient cuts for the identification of the reaction of interest. Therefore, the 
basic features of the MC simulations are discussed before details of the kinematic cuts applied 
to the data are given.  

The MC simulations were based on the Geant4 package \cite{Geant4}. All details of the detector
setup, i.e. active components and inactive materials, were implemented as precisely as known.
The quality of these simulations was already tested for other reactions analyzed from the same 
data set (see Refs.~\cite{Werthmueller_14,Oberle_14,Dieterle_15,Kaeser_16} for quasi-free production
of $\eta$ mesons, pion pairs, and $\pi\eta$ pairs from deuterium) and also for beam-time periods 
with other targets (see Refs.~\cite{Schumann_10,Zehr_12,Witthauer_13} for hydrogen and $^3$He 
targets). These analyses showed that the detector response to photon showers was correctly 
reproduced. Stringent tests came from the comparison of the results for $\eta$ photoproduction 
using the $\eta\rightarrow 2\gamma$ and $\eta\rightarrow 3\pi^0\rightarrow 6\gamma$ decays 
\cite{Witthauer_13,Werthmueller_14}. The results were in excellent agreement. Since even
small inaccuracies in photon detection efficiency would lead to significant discrepancies, this
indicates that the photon detection efficiency is well understood. The simulation of the response 
to recoil nucleons was more involved. The Geant4 package offers several different physics models 
for the strong interaction of particles with matter \cite{Apostolakis_07}. Results from 
simulations using these different models were tested against the experimental data (e.g. the 
cluster size distributions of proton and neutron hits). For protons, not much variation was 
found between the different models. For neutrons, the best agreement was achieved when the 
BERTini cascade model and the {\bf H}igh {\bf P}recision (HP) neutron model \cite{Apostolakis_07} 
were included. 

Results from the full simulation based on this model, including the electromagnetic showers of 
the photons and the recoil nucleons, are compared for several measured kinematic quantities 
in the next section. However, such simulations were not precise enough for the construction 
of the detection efficiency. Corrections derived from experimental data were necessary for 
the recoil nucleons. In particular, in the angular transition region from the CB to TAPS, inactive 
materials from support structures are complex and were not included with sufficient accuracy 
in the simulations. However, these are corrections which matter only for the exact 
values of absolute detection efficiencies for specific event topologies, but not for the 
discussion of the kinematic cuts in the next subsection. More details of the corrections 
required for the absolute normalization of cross sections are given in section \ref{sec:norm_rec}.

The input to the MC simulations was produced with event generators, which randomly generate 
events of the reactions of interest according to their kinematic characteristics. As a basis,
the event generator PLUTO \cite{PLUTO} was used, which was originally developed for heavy 
ion reactions. It had to be extended in two respects. The original version used incident particle 
beams of fixed energy. This was modified to an incident photon beam with a typical bremsstrahlung
energy spectrum. It was also not designed to describe reactions with nucleons bound 
in nuclei, so that the effects from nuclear Fermi smearing had to be implemented. 
The parameterization of the deuteron wave function in momentum space from the Paris potential 
\cite{Lacombe_81} was used. The simulated data were then analyzed with the same software 
package as the measured data. 
 
It is not sufficient to simulate only the reaction of interest. The most important
background reactions must also be simulated to optimize the cuts which discriminate 
against them. Removal of background from other reactions with the same final state, i.e. 
production of other mesons which decay to photon pairs, can be easily removed by an invariant 
mass analysis of the photon pairs. More critical are backgrounds from reactions with additional 
particles that have escaped detection. For single $\pi^0$ production on the proton, 
$\gamma p\to\pi^{0}p$, the following background contributions 
have been studied:
\begin{eqnarray}
\gamma n \to & \hspace*{-3.0cm} \pi^{0}\pi^{-}p, & \\
\gamma n \to & \hspace*{-1.5cm} \Delta^{+}\pi^{-}\to\pi^{0}\pi^{-}p, & \nonumber\\ 
\gamma p \to & \hspace*{-3.0cm} \pi^{0}\pi^{0}p, &\nonumber\\
\gamma p \to & \hspace*{0.7cm} \pi^{+}\pi^{-}\pi^{0}p, ~~~\to\eta p\to\pi^{+}\pi^{-}\pi^{0}p & \nonumber \;.
\end{eqnarray} 
Similarly, for $\pi^0$ production on the neutron, $\gamma n\to\pi^{0}n$, background from 
\begin{eqnarray}
\gamma p\to &  \hspace*{-4.5cm} \pi^{0}\pi^{+}n, & \\
\gamma p\to &  \hspace*{0.5cm} \Delta^{+}\pi^{0}\to\pi^{0}\pi^{+}n, ~~~\to\Delta^{0}\pi^{+}\to\pi^{0}\pi^{+}n, & \nonumber \\ 
\gamma n\to &  \hspace*{-4.5cm} \pi^{0}\pi^{0}n, &\nonumber\\
\gamma n\to &  \hspace*{-0.7cm} \pi^{+}\pi^{-}\pi^{0}n, ~~~\to\eta n\to\pi^{+}\pi^{-}\pi^{0}n & \nonumber 
\end{eqnarray} 
was considered. For reactions where no intermediate state is given, phase-space distributions were used. 
The $\Delta\pi$ intermediate state was explicitly included for the production of pion pairs. In the energy 
range of interest, a significant fraction of such reactions is due to sequential resonance decays 
of the type $R\to\Delta\pi\to\pi\pi N$ ($R$: any higher lying resonance) or, even more important
for charged pions, to the vertex $\gamma N\to\Delta\pi$ ($\Delta$ pion-pole or $\Delta$ Kroll-Rudermann 
like diagrams) \cite{Dieterle_15,Zehr_12}. However, the contribution from $\Delta^0\pi^0$ intermediate states
is negligible. 

All reactions were simulated for incident nucleons bound in the deuteron. The dominant background was related 
to the final states $\pi^0\pi^+ n$ and $\pi^0\pi^- p$ where the charged pion had escaped detection because 
it was emitted in the direction of the beam pipe or too low in energy. 



\subsection{Reaction identification}
\label{sub:reaction}

\begin{figure*}[!thbp]
\centerline{\resizebox{0.99\textwidth}{!}{%
  \includegraphics{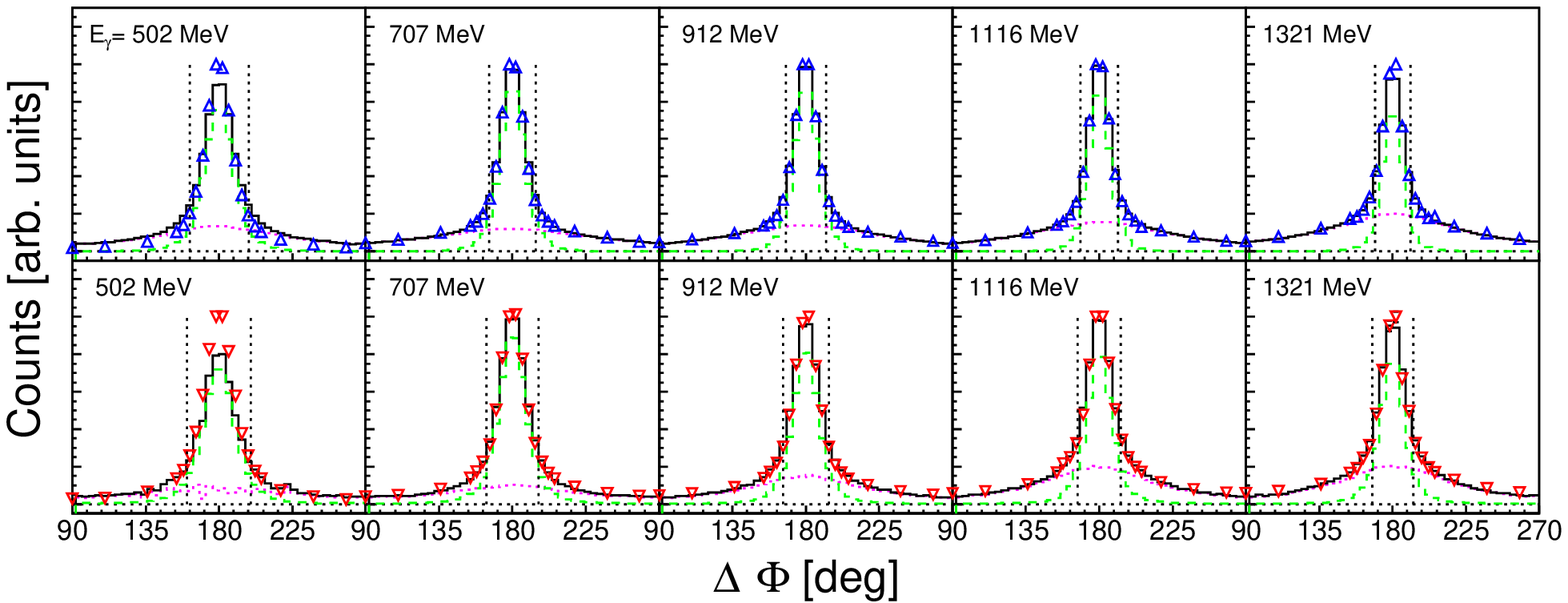}
}}
\caption{Coplanarity angle distributions for several incident photon energies for 
exclusive single $\pi^{0}$ photoproduction off the quasi-free proton (top, open upward blue 
triangles) and the quasi-free neutron (bottom, open downward red triangles) integrated over 
the full angular range. Dashed green line: MC signal, dotted magenta line: sum of 
MC background contributions, solid black line: sum of MC signal and MC background, 
dotted vertical lines: $\pm1.5\sigma$ cut positions. Spectra shown have cuts on PSA, 
a rough invariant mass cut, and a $\chi^2$ analysis for identification of recoil neutrons in CB.}
\label{fig:cop}       
\end{figure*}

\begin{figure*}[!htbp]
\centerline{\resizebox{0.99\textwidth}{!}{%
  \includegraphics{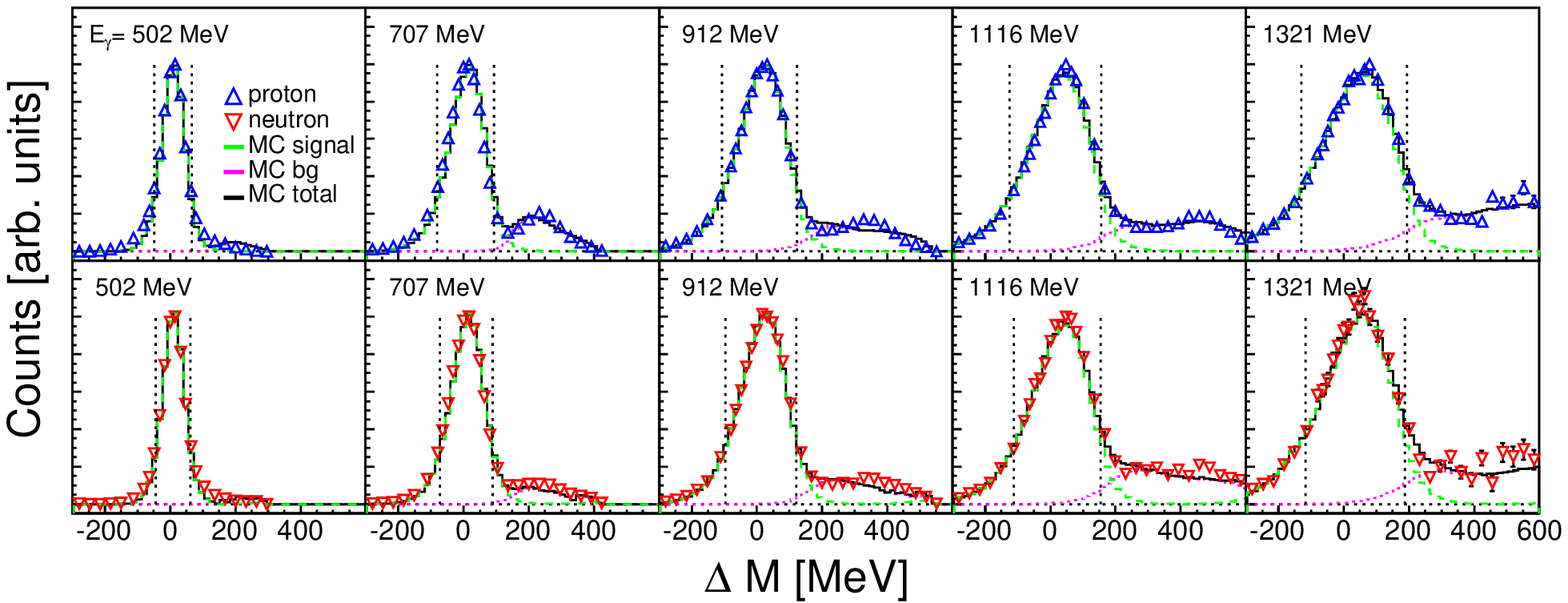}
}}
\caption{Missing mass distributions for several incident photon energies for exclusive 
single $\pi^{0}$ photoproduction off the quasi-free proton (top, open upward blue triangles) 
and the quasi-free neutron (bottom, open downward red triangles) integrated over the full 
angular range. Dashed green line: MC signal, dotted magenta line: sum of MC background 
contributions, solid black line: sum of MC signal and MC background, dotted vertical lines: 
$\pm1.5\sigma$ cut positions. Spectra with cuts as indicated in Fig.~\ref{fig:cop} and 
additionally cuts on coplanarity as indicated in Fig.~\ref{fig:cop}.}
\label{fig:mm}       
\end{figure*}
With the analysis steps discussed above, hits in the two calorimeters were tentatively assigned 
to photons, recoil protons, and recoil neutrons. Only events with exactly two photon candidates
(subsample for $\sigma_{\rm incl}$) and events with exactly two photons and a proton or a neutron 
candidate were kept for further analysis. These events were then tested for their kinematic 
characteristics to identify single $\pi^0$ production. For all kinematic observables, the measured
data were compared to the results of the MC simulations in order to test the quality of the
simulations and to estimate the size of background contributions. 

In the first step, the coplanarity of the events was analyzed. Neglecting the Fermi 
motion of the bound nucleons, there is no transverse momentum in the initial state. Consequently, 
due to momentum conservation, the reaction products, i.e. $\pi^0$ meson and recoil nucleon, must lie 
in one plane in the laboratory system. The difference $\Delta\Phi$ in azimuthal 
angle between the pion and the recoil nucleon must therefore be 180$^{\circ}$. If a further, undetected 
meson was emitted, it should deviate from this value. Due to the Fermi motion of the bound nucleons 
and the angular resolution of the detector system, this relation is broadened around the ideal value.

\begin{figure*}[!thbp]
\centerline{\resizebox{0.99\textwidth}{!}{%
  \includegraphics{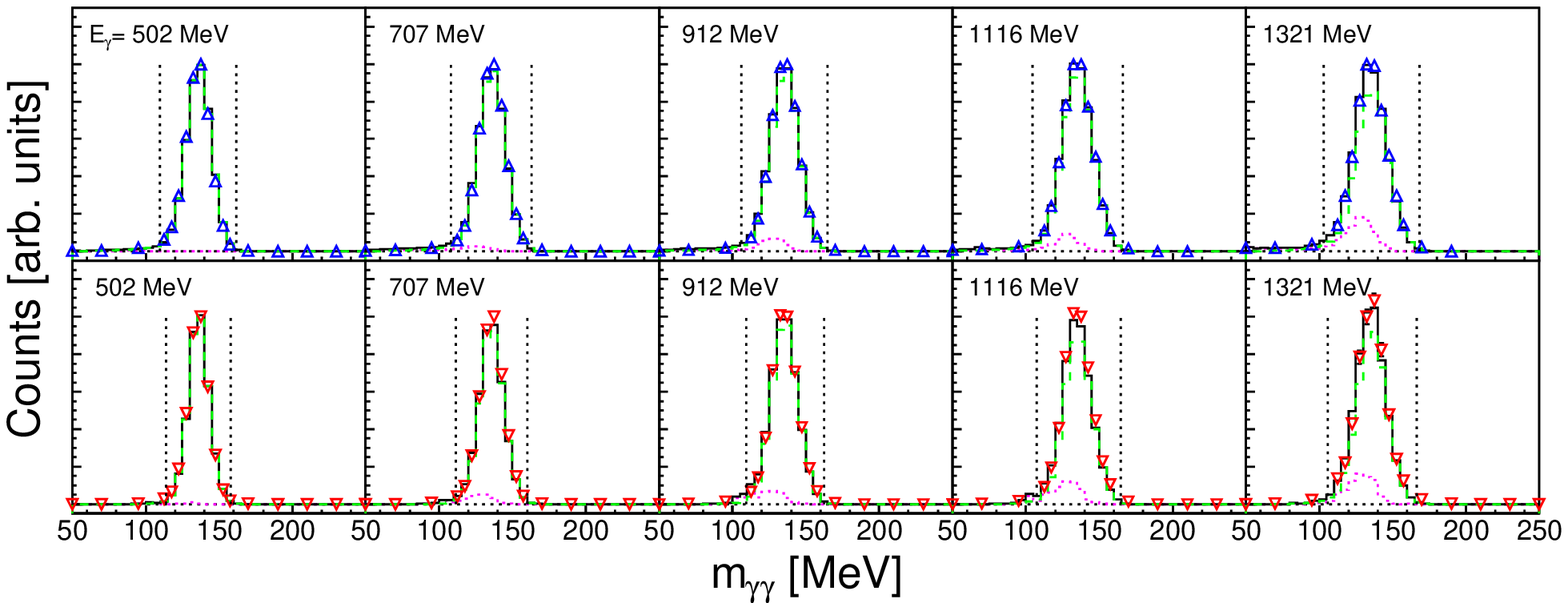}
}}
\caption{Invariant mass distributions for several incident photon energies for exclusive 
single $\pi^{0}$ photoproduction off the quasi-free proton (top, open upward blue triangles) 
and the quasi-free neutron (bottom, open downward red triangles) integrated over the full 
angular range. Dashed green line: MC signal, dotted magenta line: sum of MC background 
contributions, solid black line: sum of MC signal and MC background, dotted vertical lines: 
$\pm3\sigma$ cut positions. PSA, $\chi^2$ analysis for recoil neutrons, coplanarity, and missing 
mass cuts (as indicated in Figs.~\ref{fig:cop} and \ref{fig:mm}) were applied to the spectra.}
\label{fig:im}       
\end{figure*}

\newpage
This analysis was only possible for the exclusive reactions $\sigma_p$ and $\sigma_n$, but not
for $\sigma_{\rm incl}$, which included events without detected recoil nucleons. The results are 
shown in Fig.~\ref{fig:cop}. The experimental data were fitted with the line shapes of the simulated 
signal and background events. The background level was not high, but some components peaked at the 
position of the signal peak (although with a larger width which, in principle, would allow separation 
by a fit to these spectra). The background components were mainly due to undetected charged pions at 
extreme forward angles or small kinetic energies which did not contribute much to the transverse 
momentum balance. 
Such background is better removed by the missing mass analysis discussed below. For further 
analysis, only events within $\pm 1.5\sigma$ of the peak position were accepted (determined by 
Gaussian fits). In Fig.~\ref{fig:cop}, five examples of these spectra integrated over the 
cm-polar angle are shown. However, the actual analysis and determination of the cuts was 
dependent on incident photon energy and cm-polar angle. The good agreement between 
the measured data and the results of the MC simulations demonstrates that the detector response and 
the effects of nuclear Fermi smearing were well under control.  

For the following missing mass analysis, the recoil nucleons, if detected or not, were treated as
missing particles and their mass was reconstructed from energy/momentum conservation under the
hypothesis of single $\pi^0$ production from:  
\begin{equation}
\Delta M = \left| {\bf {P}}_{\gamma} + {\bf {P}}_{N} - {\bf {P}}_{\pi^0} \right| - m_{N}~,
\label{eq:mm}
\end{equation} 
where ${\bf {P}}_{\gamma}$, ${\bf {P}}_{N}$, and ${\bf {P}}_{\pi^0}$ are the four momenta of 
the incident photon, the incident nucleon (neglecting Fermi motion), and the final state pion, 
respectively. The mass $m_{N}$ of the participant nucleon was subtracted so that the 
missing mass $\Delta M$ should equal zero within experimental resolution and Fermi motion broadening. 
Examples, again integrated over the polar angle, are shown in Fig.~\ref{fig:mm}. Residual background 
not removed by the coplanarity cut appears at large missing masses (mainly above 200~MeV) 
and is well separated from the events from single $\pi^0$ production. 

The spectra are well reproduced by the results of the MC simulations, where the relative 
contributions of signal and background events were fitted to the data. Also, for these spectra, 
$\pm1.5 \sigma$ cuts were determined by the fits of a Gaussian distribution. These cuts are indicated 
in the figure by dotted vertical lines. The cuts at the low energy side are not necessary for the 
suppression of background. The tails at this side are due to large Fermi momenta. However, it is 
more convenient to use symmetric cuts because an asymmetric selection of Fermi momenta complicates 
further analysis.

The yields were finally extracted from the invariant mass spectra for which examples are shown in
Fig.~\ref{fig:im}. The invariant mass $m_{\gamma\gamma}$ was evaluated from:
\begin{equation}
m_{\gamma\gamma} = \sqrt{({\bf{\boldmath{P_{\gamma_1}}}} + {\bf{\boldmath{P_{\gamma_2}}}})^2} =
\sqrt{2E_{\gamma_1}E_{\gamma_2}(1-{\rm cos}(\phi_{\gamma_1,\gamma_2}))}\;,
\label{eq:invm}
\end{equation}
where ${\bf{\boldmath{P_{\gamma_1}}}}$, ${\bf{\boldmath{P_{\gamma_2}}}}$ are the four momenta of
the two $\pi^0$ decay photons, $E_{\gamma_1}$, $E_{\gamma_2}$ are their energies, and 
$\phi_{\gamma_1,\gamma_2}$ is their opening angle. These spectra were evaluated as a function of the 
incident photon energy and cm-polar angle and agreed well with MC simulations. Cuts at $\pm3\sigma$ 
were defined and are indicated in the figure. 

Residual background was quite small and corresponds to the components visible in the cut region of 
the missing mass spectra. This background was subtracted before integration of the signals. 
Altogether, agreement between experimental data and MC simulations was excellent for all investigated 
kinematic quantities, indicating that systematic effects from the analysis are small 
(see Sec.~\ref{sec:norm_rec} for a quantitative discussion).

\subsection{Reconstruction of Final-State Invariant Mass {\boldmath{$W$}}} 
\label{sec:wrecon}

The total cm energy $W$ for the photoproduction of mesons off a nucleon target, is given by:
\begin{equation}
W = \sqrt{s} = \sqrt{( {\bf {P}}_{\gamma}+{\bf {P}}_N)^2} =
\sqrt{ \left( \sum_{i=1}^{n} {\bf {P}}_{i} \right) ^2}\;,
\label{eq:w1}
\end{equation}
where ${\bf {P}}_{\gamma}$ and ${\bf {P}}_N$ are the four-momenta of the 
incident photon and the target nucleon and the ${\bf {P}}_{i}$, $i=1,...,n$ are the 
four-momenta of the final state particles (emitted mesons and recoil nucleon all in the lab frame).
For the most simple case of a free target nucleon at rest this reduces to:
\begin{equation}
W = \sqrt{2m_{N}E_{\gamma}+m_{N}^2}\;,
\label{eq:w2}
\end{equation}
with the photon beam energy $E_{\gamma}$ and the mass $m_N$ of the target. Nucleons bound in a 
nucleus are off-shell so that ${\bf {P}}_N^2\neq m_N^2$ and each fixed value of incident 
photon energy corresponds to a distribution of $W$ values, leading to the Fermi smearing of cross 
sections as a function of $E_{\gamma}$. However, this effect can be removed when $W$ is not extracted
from the incident photon energy, but from the right-hand side of Eq.~\ref{eq:w1}, using the four 
momenta of the final-state particles. The drawback of this method is that the resolution of the 
four-momenta of the final-state particles, measured with the production detector, is not 
as good as the resolution of the incident photon energy measured with the magnetic tagging 
spectrometer. 

\begin{figure}[!thbp]
\centerline{\resizebox{0.49\textwidth}{!}{%
  \includegraphics{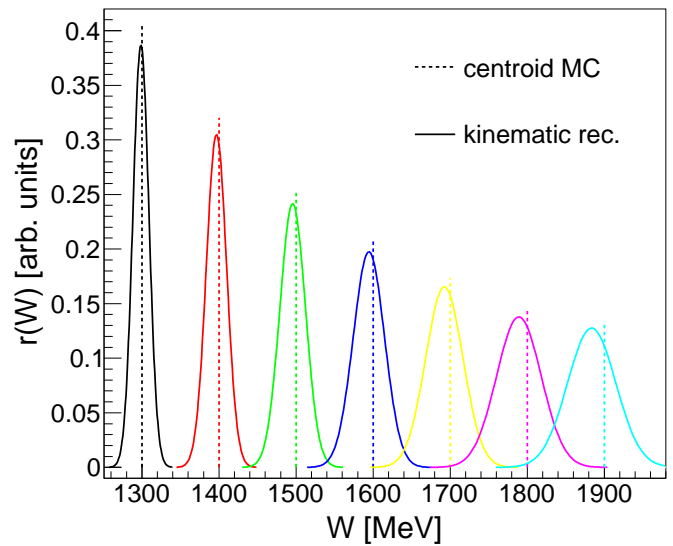}
}}
\caption{Resolution for the final state invariant mass $W$. The results of full MC simulations
of the instrumental response are shown for given values of $W$ (vertical lines).}
\label{fig:wresol}       
\end{figure}

For this reconstruction, the measured four momenta of the two decay photons were used.
There is no direct, reliable measurement of the kinetic energy of neutrons detected in the CB. 
In TAPS, in principle, time-of-flight could be used, but the resolution would not be adequate. 
However, for the reconstruction of the final state $W$ it is sufficient to measure the polar 
and azimuthal angles of the recoil nucleon. The initial state, defined by the incident photon 
of known energy and the deuteron at rest, is completely determined. In the final state, the 
four-momenta of the decay photons and the direction of momentum of the participant nucleon 
are measured. 

This means that the absolute magnitude of the momentum of the final-state recoil nucleon and the 
final-state three-momentum of the spectator nucleon are missing. These four kinematic quantities
can, however, be recovered from the four boundary conditions due to energy and momentum conservation. 
For most recoil protons the energy was directly measured by the calorimeters. However, in order 
to avoid additional systematic uncertainties in the comparison of neutron and proton cross 
sections, events with recoil protons were treated in the same way. This means that the energy 
information from the calorimeters was ignored in the reconstruction of all recoil nucleons.
 
This reconstruction also involves the determination of the polar angle of the emitted pion in the 
`true' cm system of the reaction (i.e. taking into account the momentum of the incident nucleon 
from Fermi motion). The reconstruction was done under the assumption of quasi-free production,
which means that the momenta of the incident-participant nucleon $\vec{q}_{p_i}$ and the final-state 
spectator nucleon $\vec{q}_{s_f}$ from the deuteron are related by $\vec{q}_{s_f}=-\vec{q}_{p_i}$.  

As mentioned above, the measurement of $W$ in the final state is influenced by the experimental 
resolution of the calorimeter for the photon momenta and the recoil nucleon angular resolution. 
This is shown in Fig.~\ref{fig:wresol}. The simulated response of the detector system is shown
for selected values of $W$. The relative resolution varies in the range 2 - 4\% FWHM for $W$ between 
1.3 and 1.9~GeV. Also, for the higher invariant masses, the maximum of the distributions is
slightly shifted (maximum shift: 0.9\%) with respect to the input centroid. 

\subsection{Absolute Normalization and Extraction of Cross Sections}
\label{sec:norm_rec}

The experimental yields for single pion production have been determined by integration of the
invariant mass spectra (see Fig.~\ref{fig:im} for examples) within the $\pm3\sigma$ cut ranges. 
Background from random coincidences was already removed from all spectra in 
Sec.~\ref{sub:reaction} using the coincidence condition between tagging spectrometer and 
production detector, as discussed in detail in \cite{Werthmueller_14}.

In addition, there was also background from the entrance and exit windows ($2\times 125 \mu m$ 
Kapton) of the target cells which contained `heavy' nuclei, in particular carbon. This background 
was determined with empty target measurements which were analyzed identically to the measurements 
with filled target cells. The corresponding yields, after normalization to the beam flux, were 
subtracted. Depending on the length of the target cells (4.72~cm or 3.02~cm) and on the final state 
of the reaction (with or without coincidence with recoil protons, neutrons), these 
background contributions ranged between 2 - 5\%. 

A trivial ingredient for the absolute normalization of the cross sections was the 
$\pi^0\rightarrow\gamma\gamma$ decay branching ratio taken from the 
Review of Particle Physics (RPP) \cite{PDG16} as (98.823$\pm$0.034)\%.

Furthermore, a density of 0.169~g/cm$^3$ of the liquid deuterium was used, determined with 
measurements of the target pressure. This corresponds to a surface density of 
(0.231$\pm$0.005)~nuclei/barn (4.72~cm target), and (0.147$\pm$0.003)~nuclei/barn 
(3.02~cm target), which takes into account the shapes of the convex entrance and exit windows. 

The incident photon flux was determined by a two-step measurement. The focal plane detectors 
of the tagging spectrometer were equipped with live-time gated scalers which recorded the flux 
of the scattered electrons as a function of their final-state energy. The tagging efficiency 
$\epsilon_{t}$, which is the fraction of bremsstrahlung photons which pass the collimator and 
impinge on the production target, was regularly measured at reduced beam intensity, with the 
reduction made at the electron source and no change made to the accelerator parameters.
For these measurements, a leadglass detector was moved into the primary photon beam downstream 
from the production target. Typical tagging efficiencies were in the range 60 - 70\%. Additionally, an 
ionization chamber placed downstream of the production target and just upstream of the  dump of the 
photon beam monitored the flux in arbitrary units during the production measurements. The product 
$N_{\gamma}=N_{e^-}\times\epsilon_t$ of the electron rates in the tagger and the tagging 
efficiency was taken as the incident photon flux on the target. 

\begin{figure}[!thbp]
\centerline{\resizebox{0.49\textwidth}{!}{%
  \includegraphics{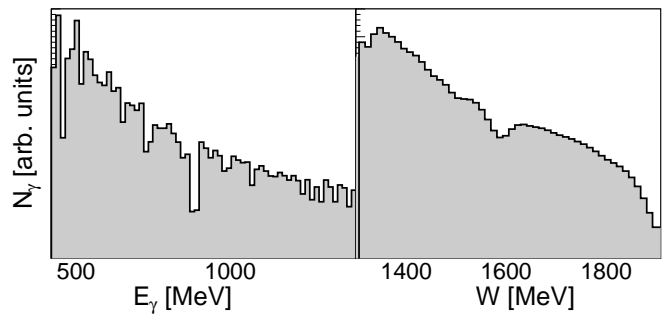}
}}
\caption{Measured photon flux for the measurement with the 3~cm target. The left-hand side shows 
the flux measured as a function of photon energy. The structures in the spectrum are due to 
tagger channels with reduced efficiency. The right-hand side shows the flux as a function of 
reconstructed $W$ after folding with the Fermi momentum distribution.}
\label{fig:flux}       
\end{figure}

An example of the flux distribution (measured with the 3~cm target) is shown in Fig.~\ref{fig:flux}. 
The original spectrum was measured as a function of the energy of the bremsstrahlung photons. 
However, for the more important analysis, in terms of the reconstructed $W$ of the final state, 
this was not the relevant quantity. The photon flux spectrum was folded with the effects of Fermi 
motion. The result is shown on the right-hand side of Fig.~\ref{fig:flux} as a function of effective $W$. 
Most of the structures from inefficient tagger channels are smeared out in this spectrum. Close to 
the upper edge of the distribution, the systematic uncertainties increase because the folding 
procedure assumes information about the photon flux at higher (untagged) photon energies.   

\begin{figure*}[thbp]
\centerline{\resizebox{1.0\textwidth}{!}{%
  \includegraphics{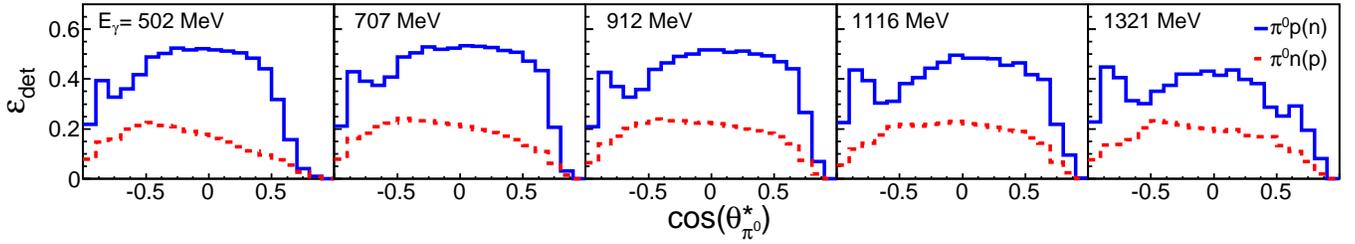}
}}
\caption{Total detection efficiency based on MC simulations and including all corrections for 
the exclusive reactions $\gamma d\rightarrow p(n)\pi^0$ (solid, blue histograms) and 
$\gamma d\rightarrow n(p)\pi^0$ (dashed, red histograms) as a function of cm angle for the same 
bins of incident photon energy as in 
Figs.~\ref{fig:cop} - \ref{fig:im}.}
\label{fig:effd}       
\end{figure*}

\begin{figure}[htbp]
\centerline{\resizebox{0.49\textwidth}{!}{%
  \includegraphics{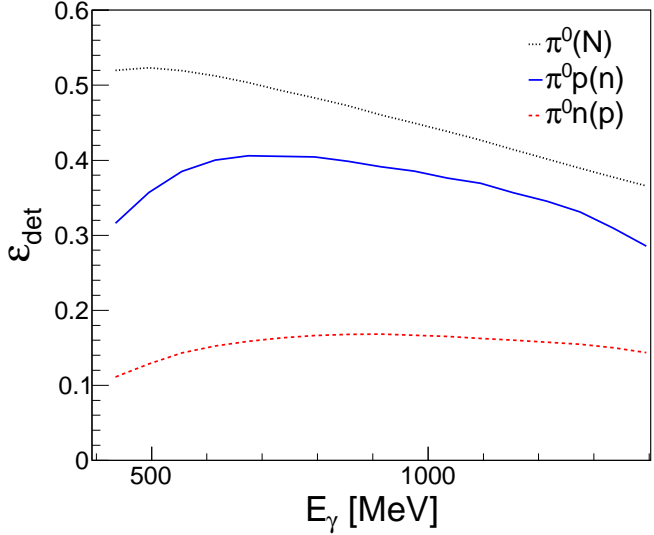}
}}
\caption{Integrated detection efficiency as a function of incident photon energy $E_{\gamma}$ 
for the inclusive reaction (dotted, black) and the exclusive reactions with detection of recoil protons 
(solid, blue) and recoil neutrons (dashed, red).}
\label{fig:efft}       
\end{figure}

The most critical ingredient for the normalization of the yields is the instrumental detection 
efficiency. The basis for this is the MC simulation discussed in Sec.~\ref{sub:MC}  
using the Geant4 code \cite{Geant4}. However, further corrections, discussed below, had to be applied. 
Examples for the detection efficiency (taking into account corrections) as a function 
of the cm polar angle and for selected bins of incident photon energy are shown for single $\pi^0$ 
production in coincidence with recoil protons and neutrons in Fig.~\ref{fig:effd}. Total detection 
efficiencies as a function of incident photon energy for these two exclusive reactions and for 
inclusive $\pi^0$ production without conditions for recoil nucleons are shown in Fig.~\ref{fig:efft}. 
The detection efficiency for recoil neutrons was roughly in the 30\% range, while recoil protons were 
detected with efficiency above 90\%. The structure in the angular dependence of the detection efficiency 
for recoil protons is due to the transition region between CB and TAPS. This effect was less important 
for recoil neutrons, which are not affected so much by inactive materials. The detection efficiency at 
extreme pion-forward angles was very low, so that no results for pion-polar angles larger than 
cos($\theta^{\star}_{\pi})>0.9$ were obtained. This was caused by the experimental trigger conditions 
discussed below.

The agreement between the experimental results and the output from the MC simulations, as far 
as the shapes of the distributions of kinematic observables such as coplanarity, missing mass, 
and invariant mass discussed in Sec.~\ref{sub:reaction} are concerned, is excellent. However, 
there are two issues which required more detailed investigation. 

The first arises from the hardware thresholds used in the experiment trigger and for the readout
of the detector elements. The NaI modules of the CB detector were equipped with two leading
edge (LED) discriminators per crystal and the modules of the TAPS detector with an LED and
a CFD (constant fraction discriminator) per crystal. The first discriminator system was used for
trigger purposes and the second (in case of TAPS, the CFDs) for the readout pattern of the 
detector. 

For the trigger, as discussed in Sec.~\ref{sec:setup}, CB and TAPS were subdivided into 
logical sectors. If the signal from at least one crystal in a sector exceeded a threshold 
($\approx$30~MeV in CB, $\approx$35~MeV in TAPS) that sector contributed to the event multiplicity 
which was two for the measurements discussed here. For events which satisfied the trigger condition,  
the second discriminator system with much lower thresholds (2~MeV for CB and 3-4~MeV for TAPS) 
generated the pattern of activated crystals from which energy and timing information was processed 
and stored. The discriminator thresholds were calibrated with the measured data and software thresholds 
above the maximum hardware thresholds were applied to experimental data and MC simulations in order 
to have well defined conditions.  

\begin{figure}[htbp]
\centerline{\resizebox{0.49\textwidth}{!}{%
  \includegraphics{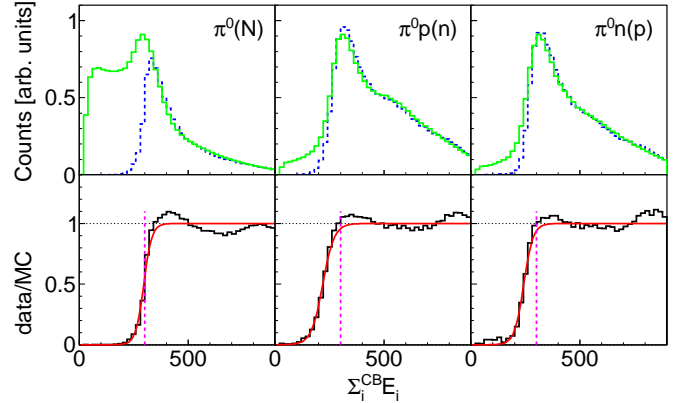}
}}
\caption{Determination of the CB energy sum threshold. 
Upper row: raw count rates. Dashed (blue): experimental data, slid (green): MC simulation. 
Lower row: ratio of experimental data and simulation (black histogram). Smooth (red) curves:
fit to data (see text). Vertical lines: preset hardware threshold. 
Both rows for inclusive data and data in coincidence with recoil protons and neutrons.}
\label{fig:sthres}       
\end{figure}

More involved was the implementation of the CB sum-threshold trigger in the simulations.
This trigger was efficient for the selection of hadronic events and significantly reduced the count 
rate from electromagnetic background. It was set such that only events with a total energy
deposition of roughly 300~MeV in the CB were accepted. However, there were several systematic 
difficulties with it. A trivial one was that the energy deposition of recoil neutrons in
the calorimeter is basically random. Depending on whether and where neutrons induce hadronic 
reactions they can deposit very different amounts of energy and there is no correspondence between
their kinetic energy and the energy they deposit in the calorimeter. To address this problem, events 
from the experimental data and also from the MC simulations were only accepted when the photon 
hits in the CB alone exceeded the sum threshold. Events where the recoil nucleon had to contribute 
to the sum trigger condition were discarded. This was also done for recoil protons in order to 
avoid systematic uncertainty in the comparison of proton and neutron data.    

The sum-threshold trigger acted on the electronically generated analog sum of the uncalibrated 
output-voltage signals from the CB detector modules. The HV for the individual 
modules was set in a way that the deposited energy to output-voltage relation was similar for 
all crystals, but this was only an approximation. Therefore, the implementation of this trigger 
condition into the MC simulations required a detailed analysis. In the first step, the data were 
analyzed with a high software threshold for the analog sum (400~MeV instead of the nominal
300~MeV of the experiment) to make sure that all simulated events that pass this threshold would 
have also passed the hardware threshold. This gave a reasonable approximation of the energy and 
angular dependence of the cross section as input for further simulations of the effect of the
hardware trigger. For the correct software implementation of the sum trigger, the experimental data 
and the results of the MC simulations had to be `de-calibrated' because the hardware threshold acted 
on the sum of uncalibrated output voltages. Otherwise, the contribution of individual modules to 
the sum energy would have been over- or under-estimated, depending on their calibration constants.

Fig.~\ref{fig:sthres} shows the experimental and simulated distributions of the CB sum energy 
for inclusive and exclusive reactions (upper row) and their ratio (lower row), where no energy sum 
threshold was applied in the simulations. The preset hardware energy threshold of 300~MeV is indicated 
in the lower row by the vertical lines. 

The ratio was fitted by a cumulative distribution function of the type (red curves in 
Fig.~\ref{fig:sthres}):
\begin{equation}
f(E^{CB}) = \frac{A}{ 1+{\rm exp}\left (\frac{\bar{E}-E^{CB}}{B} \right) }\;,
\end{equation}
where $A$, $B$, and $\bar{E}$ are free parameters, the latter corresponding approximately to 
the applied hardware threshold. For the final simulation of detection efficiencies, MC events 
in the region where $f(E^{CB})$ was zero were discarded, events where $f(E^{CB})=1$ were accepted, 
and events in the transition region were weighted with $f(E^{CB})$.   

The second complication was due to the detection of the recoil nucleons. Protons and neutrons with 
relatively low kinematic energies were critical. Special packages for low energy nucleons were 
used in the MC simulations but, particularly in the transition region between CB and TAPS, this 
was not good enough. The material budget in the transition region between the CB and TAPS (inactive 
materials from support structures and cables) was not represented with sufficient accuracy in the MC 
simulations.
 
The resulting effects were negligible for photons, small for recoil neutrons, but significant 
for recoil protons. However, one should note that the simulation of neutron detection efficiencies
is in general more involved than for protons. Therefore, detection efficiencies for recoil 
nucleons were cross checked with experimental data from measurements with a liquid hydrogen target. 
The reactions $\gamma p\rightarrow p\eta$ and $\gamma p\rightarrow p\pi^0\pi^0$ were analyzed for 
the detection efficiency of recoil protons and the reaction $\gamma p\rightarrow n\pi^0\pi^+$ for 
the detection efficiency of recoil neutrons. Single $\pi^0$ production off the proton could not 
be used because the hydrogen data were measured with a multiplicity-three trigger (for $\eta$
production the $\eta\rightarrow 6\gamma$ decay was used). In both cases, the detection efficiency 
was model-independently extracted from the yields of the respective meson production reactions 
with and without detection of the recoil nucleons. A matrix of detection efficiency
as a function of laboratory nucleon kinematic energies and polar angles was built. The same matrix 
was constructed for the MC simulations of the reactions from the free proton target. The ratio of 
these two distributions was then used to correct the simulated recoil nucleon detection efficiencies 
for the deuterium target. Typical corrections were below the $\pm 10\%$ level. 

The results from the two beam times using the 4.72~cm target (140~h of beam time) and the 3.02~cm 
target (190~h), which had comparable statistical quality, were in excellent agreement and were
averaged.   

\subsection{Systematic Uncertainties}
\label{sec:syseff}
Global systematic uncertainties arose from the absolute normalization due to the target surface 
density and the incident photon flux. Also in this category was the uncertainty due to the 
subtraction of the contribution from the target-cell windows. These uncertainties were neither 
energy nor angle dependent (the empty target distribution might have been so, but was so small 
that this could not be investigated). They were estimated at 3\% for the photon flux, 4\% for 
target density (mainly due to uncontrolled deformations of the target cell in the cooled state), 
and 2.5\% for the empty target subtraction (which is 50\% of the total empty target yields 
and probably overestimated). The total overall uncertainty was estimated at 7\%. 
This overall uncertainty is not included in the systematic uncertainty bands shown in the 
figures of the results section \ref{sec:results}.  

More important were the energy and angle dependent uncertainties from trigger conditions,
analysis cuts, and MC simulations. They were estimated by varying the cut conditions in the 
analysis and by artificially replacing the hardware thresholds by higher software thresholds 
(e.g. the CB energy-sum threshold from 300~MeV to 400~MeV). The empirical corrections to
the recoil nucleon detection efficiencies were also taken into account. 

Typical systematic uncertainties from these sources were around 5\% for incident photon energies above 
700~MeV and rose to about 15\% for photon energies around 500~MeV. The largest systematic uncertainties 
arose at extreme forward and backward pion angles, in particular for low incident 
photon energies. This is mainly due to the CB sum-energy trigger. Decay photons from pions close to 
polar angles of 0$^{\circ}$ or 180$^{\circ}$ degrees were not likely to hit the CB. Therefore, only 
few events from very asymmetric decays of the pion triggered the sum threshold, which made this class 
of events prone to systematic effects from details of the hardware thresholds. Events at extreme 
pion-forward angles (cos($\theta_{\pi}^{\star})>0.9$) could not be analyzed because for such events, most 
decay photons were outside the angular range of the CB so that the sum threshold did not trigger.

\subsection{Correction of Final State Interaction Effects}
\label{sec:FSI}

The production of mesons from quasi-free nucleons bound in a nucleus is also influenced by 
final-state interactions. For the special case of pion production from the deuteron, such 
interactions may arise in the final state $NN$ system and/or the $\pi N_s$ system 
($N_s$: spectator nucleon). $\pi N_p$ rescattering ($N_p$: participant nucleon) also
contributes for reactions on a free proton target. The magnitude and the energy and angular dependence 
of FSI can differ significantly between reactions. Previous experiments have shown that FSI for 
$\eta$ photoproduction off deuterons in the energy range discussed here is negligible for cross 
sections and also for polarization observables 
\cite{Jaegle_11,Werthmueller_13,Werthmueller_14,Witthauer_16,Witthauer_17,Witthauer_17a}. 
Also for photoproduction of $\eta '$ mesons, no significant effects were observed \cite{Jaegle_11a}. 
In the production of pion- and $\pi\eta$-pairs, FSI was significant but moderate (typically in the
10 - 20\% range, up to 30\% for $\pi^0\eta$ pairs) \cite{Oberle_13,Oberle_14,Dieterle_15,Kaeser_16,Kaeser_15}. 
Important FSI effects were also observed for the production of charged pions in the 
$\gamma d\rightarrow pp\pi^-$ reaction \cite{Tarasov_11,Chen_12}. 

The present results for photoproduction of $\pi^0$ mesons show large deviations 
(see Sec.~\ref{sec:results}) between the results for free and quasi-free protons bound in the 
deuteron. Most deviations are in the absolute scale of the cross section, while, apart from extreme 
forward angles, the shape of the angular distributions is not much affected. This observation is 
supported by the measurement of the helicity components of the total cross section: 
$\sigma_{3/2}$ (parallel photon and nucleon spin) and $\sigma_{1/2}$ (antiparallel spins) 
\cite{Dieterle_17}. The ratio of the $\sigma_{1/2}$ and $\sigma_{3/2}$ components is almost identical 
for free and quasi-free protons, with only the absolute scale of both cross sections modified. 

For reactions with pions emitted at extreme forward angles, most of the momentum of the incident 
photon is transferred to the pion and the relative momentum between `participant' and `spectator' 
nucleons is small, giving rise to large $NN$ FSI. This happens also for $\eta$ and $\eta '$ production. 
However, in contrast to pion production, those reactions are dominated by the $E_{0+}$ multipole 
from the excitation of $S_{11}$ nucleon resonances. This reaction mechanism requires a spin-flip 
of the participant nucleon so that the two nucleons have antiparallel spin in the final state, 
while for pion production the deuteron-like configuration with parallel spins is more important, 
giving rise to very different $NN$ FSI. 

A model analysis of FSI for $\pi^0$ production off the deuteron \cite{Tarasov_16} predicts that it
is only significantly different for participant protons and neutrons at extreme forward pion angles 
(for which we do not have data). However, the absolute predicted scale of the effects for the proton 
target was not in quantitative agreement with observations, so that these predictions could not be used 
to correct the neutron data for FSI. Further modeling is under way \cite{Nakamura_17}, but there are 
not yet final results.

Currently, the only reasonable correction of the quasi-free neutron results for FSI 
assumes that it is similar for protons and neutrons bound in the deuteron. 
For protons it can be determined experimentally by a comparison of the reactions on free 
and quasi-free protons. The ratio of these proton cross sections can then be used to correct
the quasi-free neutron cross section:
\begin{equation}
\frac{d\sigma^f_n}{d\Omega}(z,W)=
\frac{d\sigma^{qf}_n}{d\Omega}(z,W)\times
\frac{<d\sigma^f_p>}{d\sigma^{qf}_p}(z,W)\;,
\label{eq:fsi_corr}
\end{equation} 
with $z={\rm cos}(\theta_{\pi}^{\star})$ and the subscripts $p$ and $n$ denote proton and 
neutron cross sections and the superscripts $f$ and $qf$ free and quasi-free cross sections.

However, one cannot directly compare measured quasi-free and free proton cross sections. 
The energy resolution for the quasi-free proton data includes the effects from the kinematic 
reconstruction of $W$ for the final state, while $W$ is directly taken from the incident photon 
energy measured with the tagging spectrometer for the free proton data. Due to this effect, 
structures such as the resonance bumps in the photoproduction of pions appear `dampened' for the 
quasi-free reaction and the ratio of free to quasi-free data develops artificial structures. 
Therefore, the measured free proton cross section $d\sigma^f_p/d\Omega(z,W)$ was not used in 
Eq.~\ref{eq:fsi_corr}. Instead this cross section was folded with the experimental resolution of 
the $W$ reconstruction of the quasi-free measurement. The result of the folding is denoted
by $<d\sigma^f_p>/d\Omega(z,W)$. This avoids artificial structures, but does not correct 
the finite resolution effects.

An advantageous side-effect of this FSI correction for the neutron cross section is that systematic
uncertainties from this experiment (hardware thresholds, overall normalization, MC simulations 
of photon showers, etc.) cancel in Eq.~\ref{eq:fsi_corr} in the $d\sigma^{qf}_n/d\sigma^{qf}_p$ ratio
(except those arising from the proton and neutron detection efficiencies). 

For all results shown in the next section it is mentioned in the figure captions when data have been 
corrected for FSI effects as described above. All other results are uncorrected quasi-free data.

\section{Results}
\label{sec:results}

\begin{figure*}[!thbp]
\centerline{\resizebox{0.99\textwidth}{!}{%
  \includegraphics{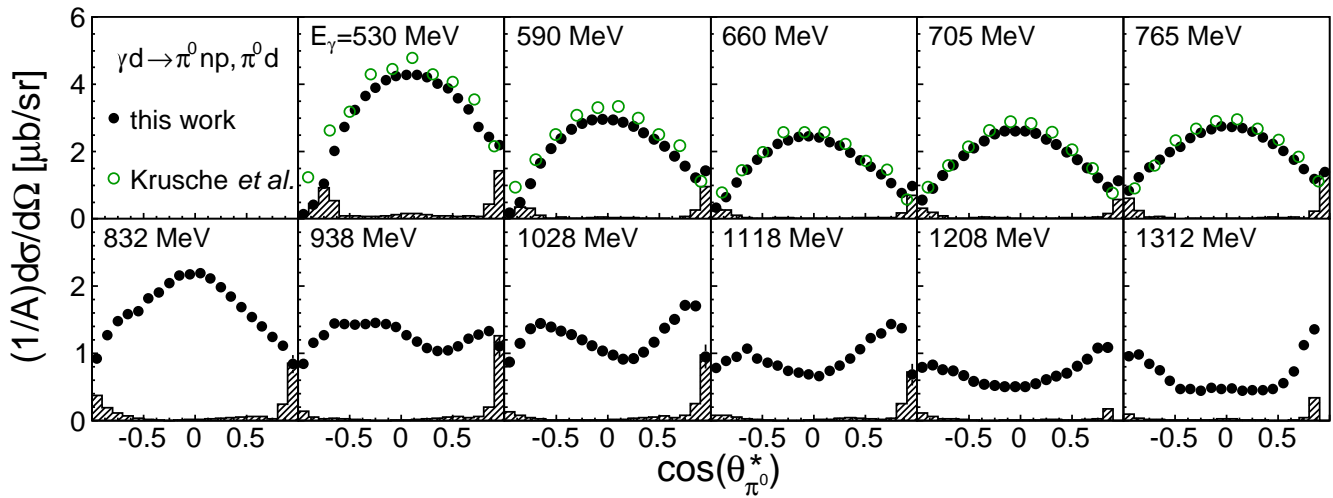}
}}
\caption{Selected differential cross sections as function of the incident photon energy for 
quasi-free inclusive single $\pi^{0}$ photoproduction compared to former results 
\cite{Krusche_99}. Full black circles: Present results, open green circles: results from 
\cite{Krusche_99}. Cross sections normalized by $A$=2, the number of nucleons (i.e. average
nucleon cross section). Shaded bands: systematic uncertainty excluding 7\% overall normalization 
uncertainty.}
\label{fig:inclusive_old}       
\end{figure*}

\begin{figure*}[!thbp]
\centerline{\resizebox{0.85\textwidth}{!}{%
  \includegraphics{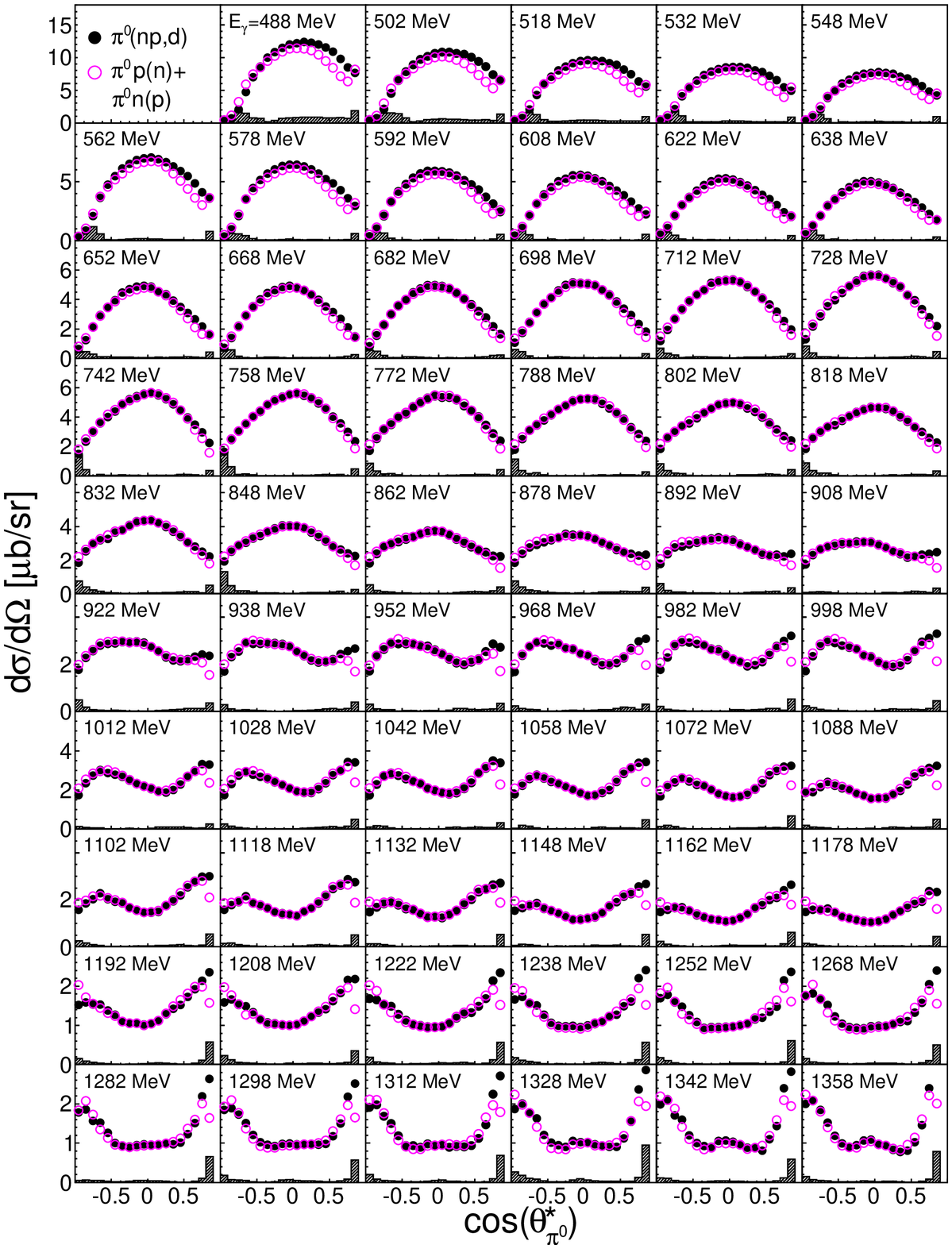}
}}
\caption{Differential cross sections as a function of the cm polar angle for different bins of
incident photon energy $E_{\gamma}$ (central values of the bins are labeled in the figures).
Black, filled dots correspond to the inclusive cross section $d\sigma_{\rm incl}/d\Omega$,
including all single $\pi^0$ production reactions with a (np) or d final nucleon state.
Magenta circles show the sum $d\sigma_{p}/d\Omega$+$d\sigma_{n}/d\Omega$ of the exclusive cross
sections in coincidence with recoil protons and neutrons. The black histograms indicate the 
systematic uncertainty of the inclusive cross section (without the 7\% overall normalization
uncertainty).
}
\label{fig:dcs_inc}       
\end{figure*}

First, we discuss the results for the inclusive cross section $\sigma_{\rm incl}$.  
The only condition for such events was the identification of a $\pi^0$ meson and
the exclusion of the production of further mesons by the missing mass analysis. An additional
charged or neutral hit (due to recoil neutrons, recoil protons, or recoil deuterons) was 
accepted, but not required. This analysis was more prone to background than 
the exclusive analyses discussed below because coplanarity conditions could not be used. 
Also the kinematic reconstruction of the final state was not possible because a significant 
fraction of events, detected without a recoil nucleon, were kinematically under determined
so that only the incident photon energy, measured by the tagging spectrometer, was available.

\begin{figure*}[!thbp]
\centerline{\resizebox{0.98\textwidth}{!}{%
  \includegraphics{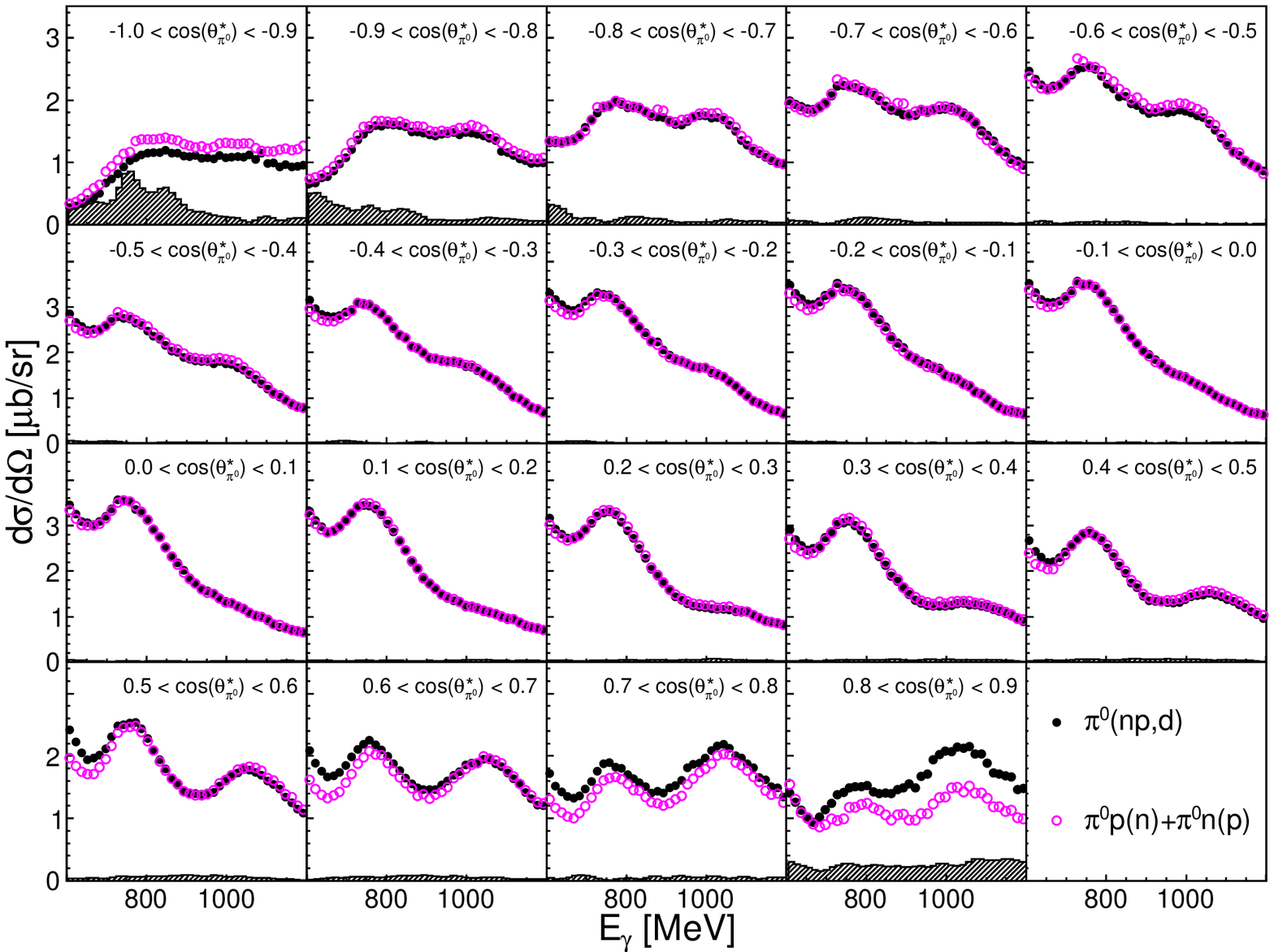}
}}
\caption{Differential cross sections for the inclusive reaction $\gamma d\rightarrow \pi^0 X$
(black dots) and sum of exclusive cross sections (open magenta circles) as a function of the 
incident photon energy for different cm polar angle bins. Notation as in Fig.~\ref{fig:dcs_inc}.}
\label{fig:dcs_inc_e}       
\end{figure*}

Several aspects of the results from the inclusive reaction, not discussed in the preceding
letter \cite{Dieterle_14}, are interesting. First of all, these are the only results from
the present experiment which can be compared to previous data. In Fig.~\ref{fig:inclusive_old}, 
the present results for some typical energy bins are shown and compared to previous results from 
\cite{Krusche_99}. For the energy ranges where previous measurements are available, agreement of the shape 
of the angular distributions is excellent. The two results differ on an absolute scale
by up to 10\%. The overall normalization uncertainty for the two experiments is almost equal
(7\% for the present and 6\% for the previous data \cite{Krusche_99}) so that no scale can be preferred.  
The agreement is not trivial because the instrumental detection efficiency (solid angle coverage) 
was very different for the two experiments ($\approx 25\%$ of the full solid angle for \cite{Krusche_99} 
and $\approx 93\%$ of $4\pi$ for the present results). This corresponds to more than an order of magnitude 
in the detection efficiency for photon pairs. Also, the determination of the detection efficiency 
was done in different ways for the two experiments. For the results in \cite{Krusche_99}, 
the detection efficiency was simulated in bins of laboratory polar angle and laboratory 
kinetic energy of the pions, while an event generator taking into account the roughly known 
angular distributions and effects of Fermi motion was used for the present results. 
Systematic uncertainties for these two approaches come from different sources. Results 
from earlier measurements with untagged photon beams and without discrimination against 
production of pion pairs are not shown; references can be found in \cite{Krusche_99}.

Furthermore, a comparison of the results for the inclusive reaction and the exclusive 
reactions, in coincidence with recoil protons and recoil neutrons, provides stringent boundaries 
on systematic uncertainties for the detection of recoil protons and recoil neutrons. 
The results for the inclusive reaction and the sum of the exclusive reactions are compared 
in Fig.~\ref{fig:dcs_inc} (angular distributions) and in Fig.~\ref{fig:dcs_inc_e} 
(excitation functions in bins of cm-polar angle). Apart from the extreme forward and backward angles 
(discussed below), the agreement between the two data sets is excellent. The inclusive cross 
section $\sigma_{\rm incl}$ depends only on the detection efficiency of the $\pi^0$-decay photons. 
The exclusive cross sections $\sigma_p$, $\sigma_n$ also depend on the very different 
detection efficiencies of recoil protons ($>90\%$) and recoil neutrons ($\approx 20 - 30\%$). 
Therefore, the good agreement between the two analyses means that the recoil nucleon 
detection efficiencies are well under control. Similar results have previously been found for other 
reactions analyzed from the same data sample ($\eta$ production \cite{Werthmueller_14}, 
photoproduction of $\pi^0$ pairs \cite{Dieterle_15}, and of $\eta\pi$ pairs \cite{Kaeser_16}). 
This is evidence that the detection of recoil nucleons is understood. 

\begin{figure*}[!thbp]
\centerline{\resizebox{0.85\textwidth}{!}{%
  \includegraphics{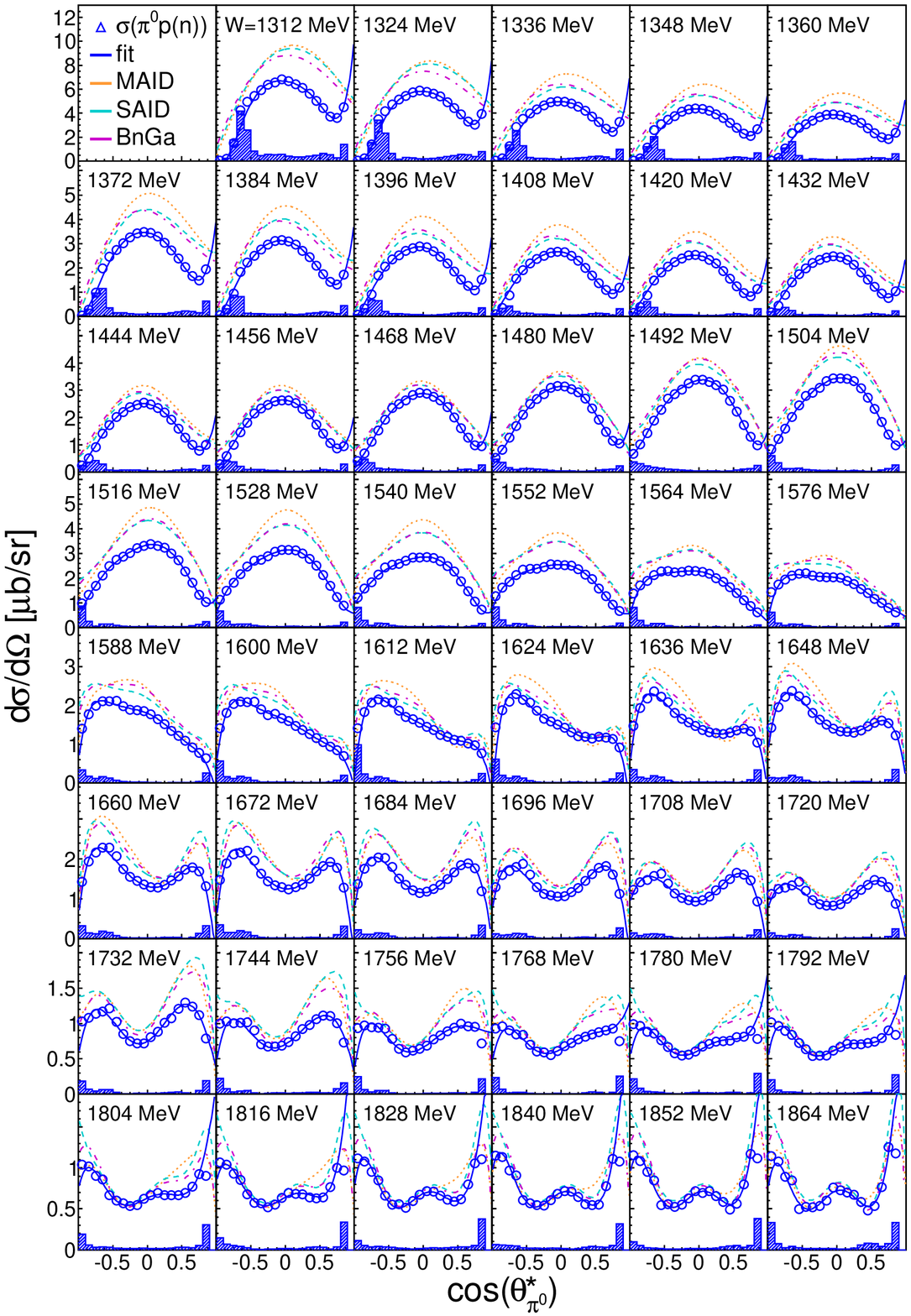}  
}}
\caption{Differential cross sections for exclusive single $\pi^{0}$ photoproduction off the 
quasi-free proton. Open blue circles: experimental data, histograms: systematic uncertainty, 
solid blue lines: Legendre fit to measured cross sections, model results: dashed cyan line: SAID, 
dotted orange line: MAID, dash-dotted magenta line: BnGa.
}
\label{fig:dcsw_p}       
\end{figure*}

\begin{figure*}[!thbp]
\centerline{\resizebox{0.85\textwidth}{!}{%
  \includegraphics{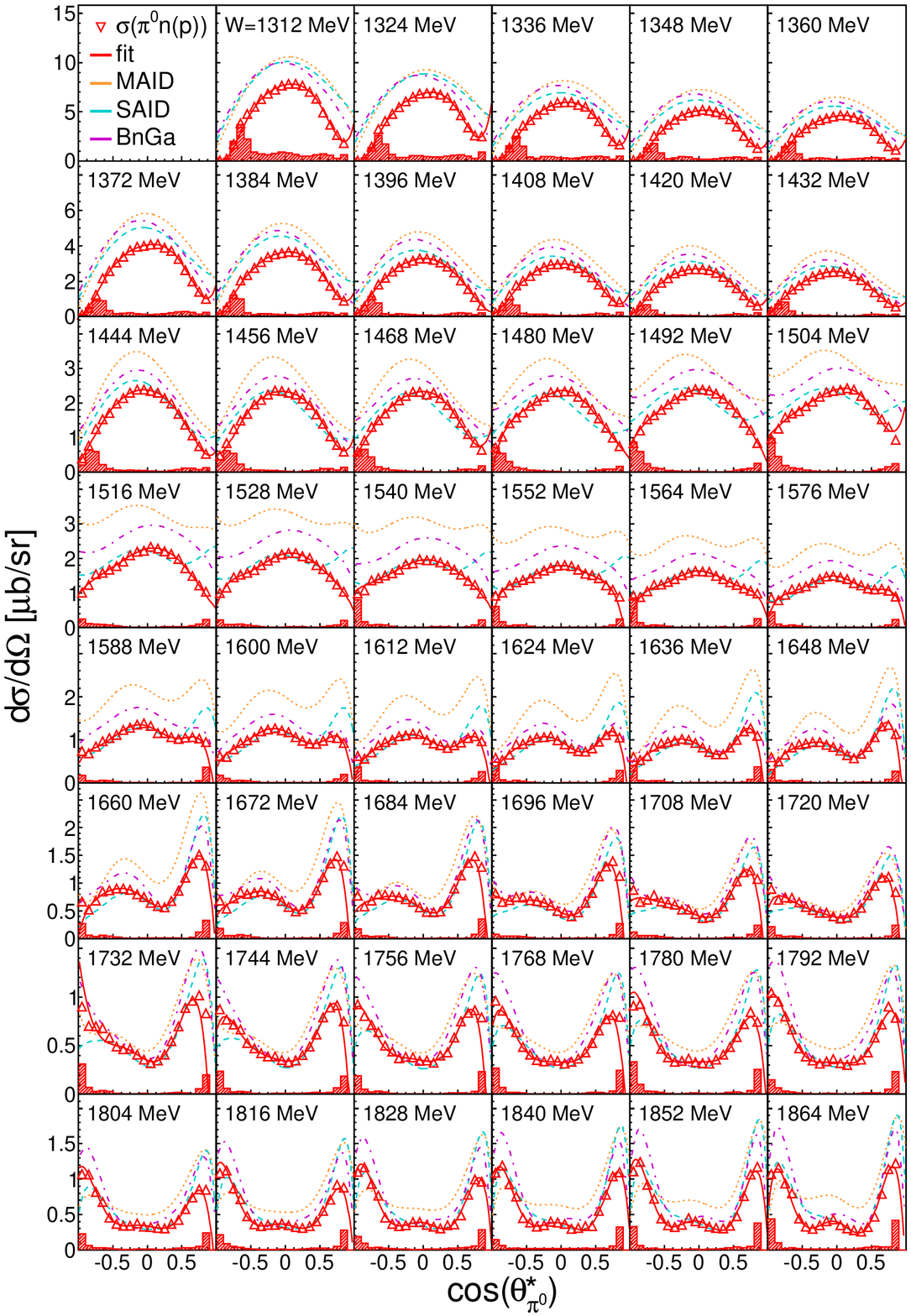}  
}}
\caption{Differential cross sections for exclusive single $\pi^{0}$ photoproduction off the 
quasi-free neutron. Open red triangles: experimental data, histograms: systematic uncertainties, 
solid red lines: Legendre fit to measured cross sections, model results: dashed cyan line: SAID, 
dotted orange line: MAID, dash-dotted magenta line: BnGa.
}
\label{fig:dcsw_n}       
\end{figure*}

\begin{figure*}[!thbp]
\centerline{\resizebox{0.85\textwidth}{!}{%
  \includegraphics{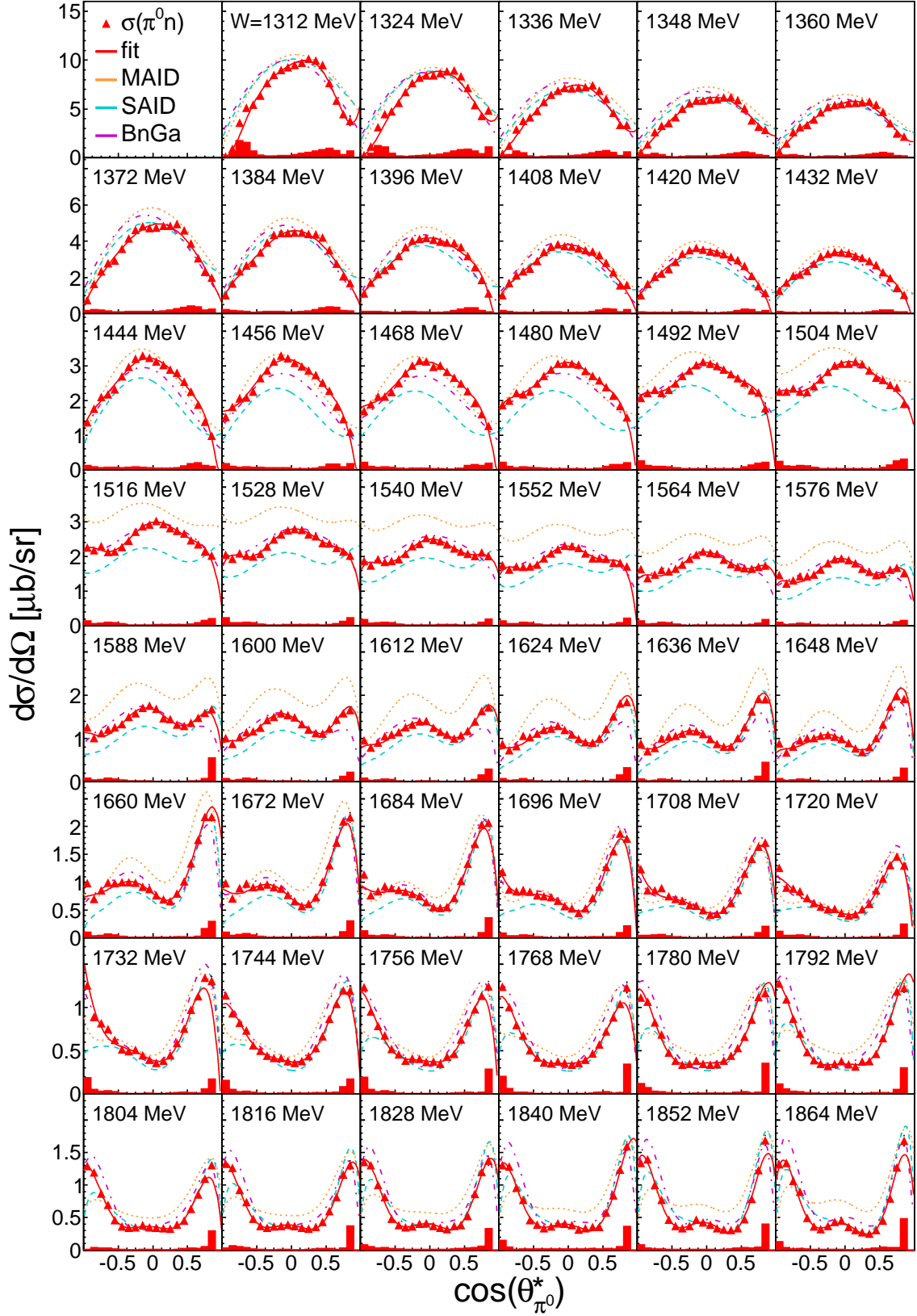}  
}}
\caption{Differential cross sections for exclusive single $\pi^{0}$ photoproduction off the 
free neutron (full red triangles). These are quasi-free data corrected for FSI effects.
Histograms: systematic uncertainties, Red solid lines: Legendre fit to measured data, 
model results: dashed cyan lines (SAID), dotted 
orange lines (MAID), dash-dotted magenta lines (BnGa).
}
\label{fig:dcsw_corr_n}       
\end{figure*}

\begin{figure*}[!thbp]
\centerline{\resizebox{0.99\textwidth}{!}{%
  \includegraphics{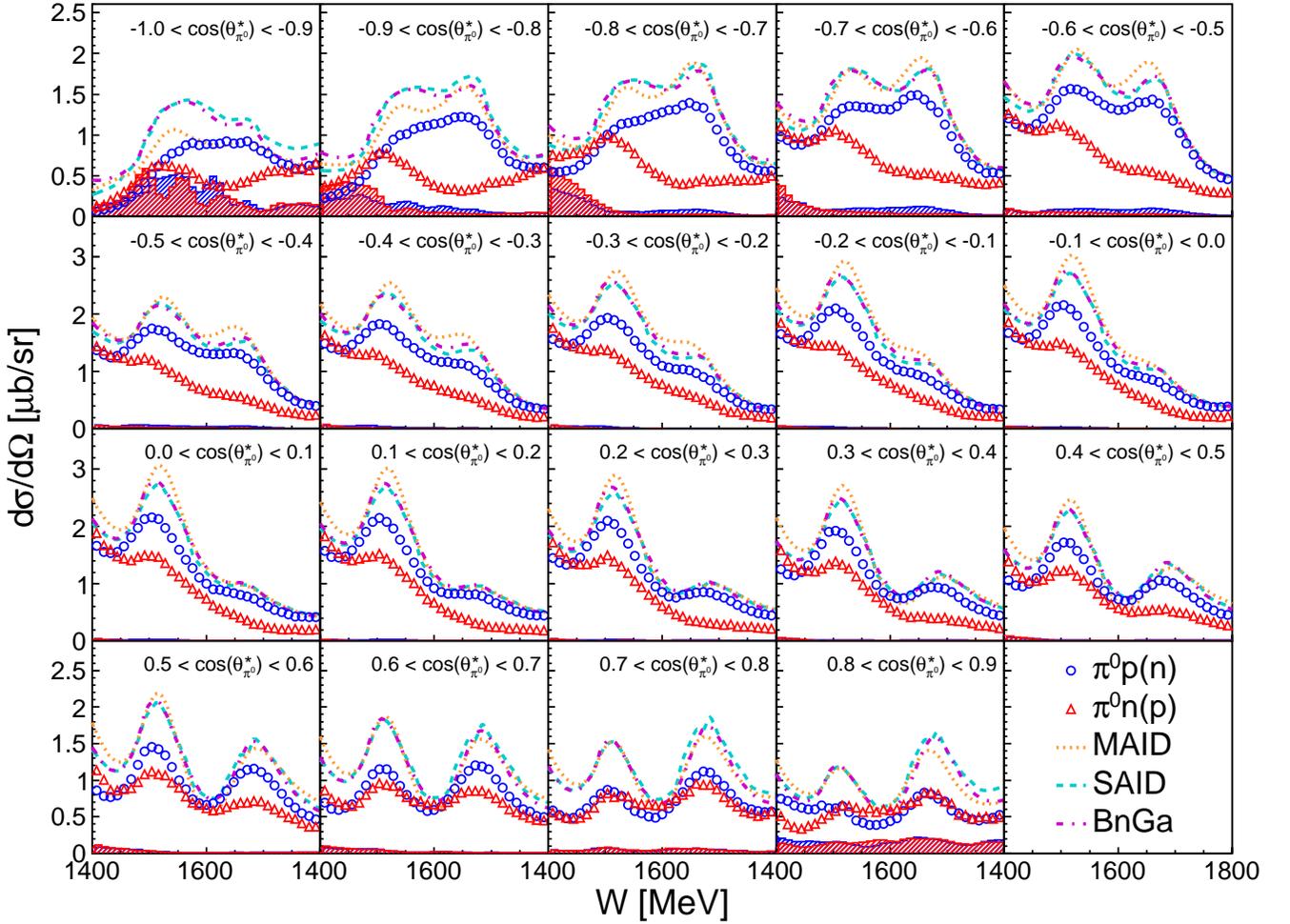}  
}}
\caption{Differential cross sections as a function of the final state invariant mass for 
exclusive single $\pi^{0}$ photoproduction off the quasi-free proton (blue, open circles)
and the quasi-free neutron (red, open triangles).
Histograms: systematic uncertainties. Lines: model results for the free proton with notation
as in Fig.~\ref{fig:dcsw_p}.
}
\label{fig:dcsw_e_np}       
\end{figure*}

\begin{figure*}[!htbp]
\centerline{\resizebox{0.99\textwidth}{!}{%
  \includegraphics{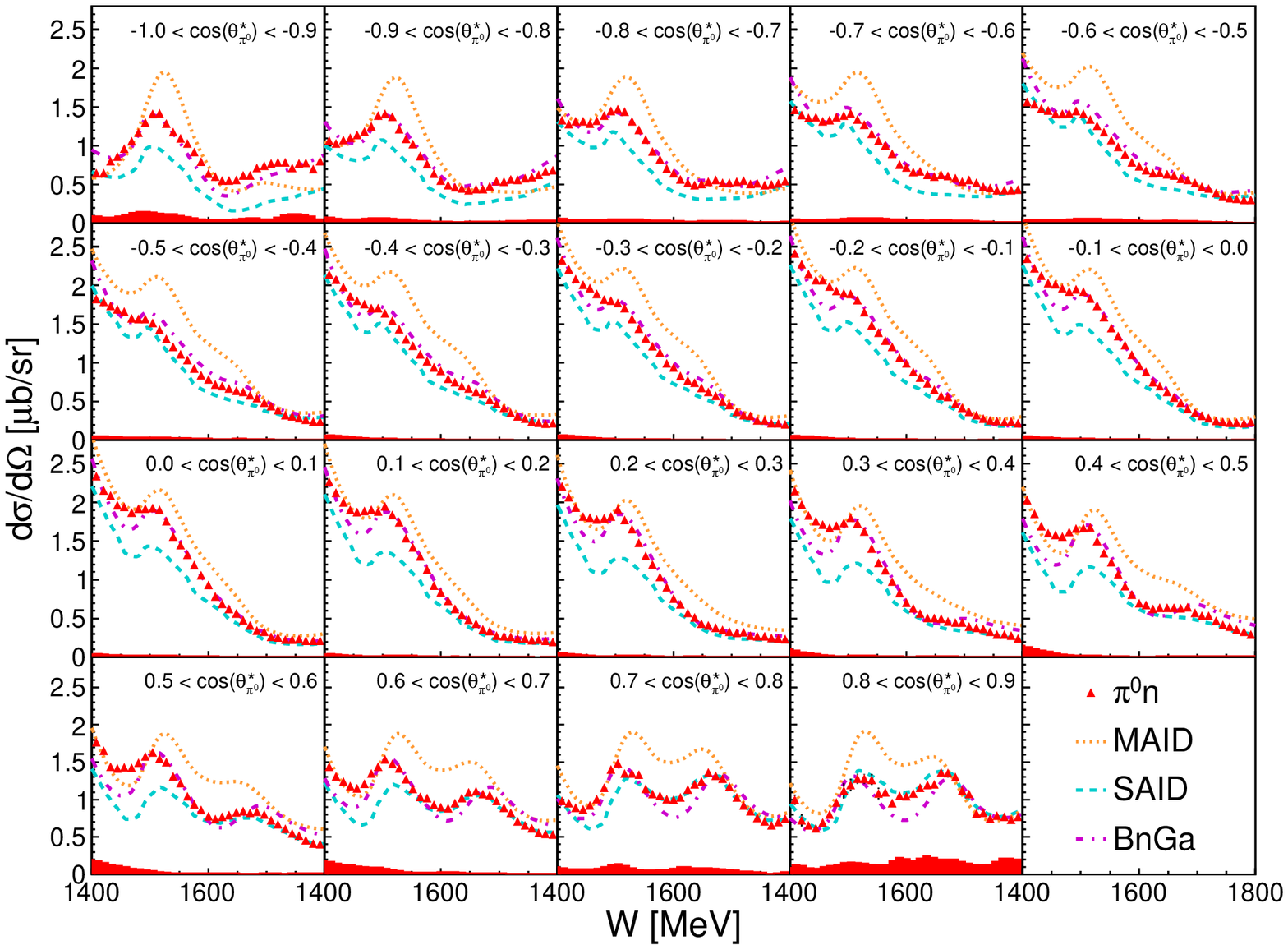}  
}}
\caption{Differential cross sections as a function of the final state invariant mass for 
exclusive single $\pi^{0}$ photoproduction off the free neutron (i.e. quasi-free neutron data
with correction of FSI effects). 
Red triangles: experimental data, histograms: systematic uncertainties. 
Notation for model results as indicated in Fig.~\ref{fig:dcsw_corr_n}.
}
\label{fig:dcsw_e_ncorr}       
\end{figure*}

The deviations at extreme pion backward angles are within the quoted systematic uncertainties,
which are mostly due to the sum-threshold trigger. However, this effect should be similar
for the inclusive cross section and the sum of the exclusive cross sections because in both 
cases, only photons were accepted in the software trigger. Therefore, the quoted systematic 
uncertainty certainly overestimates the relative systematic uncertainty between the two results, 
but it should be considered when either result is compared to other data or model results. 
For the exclusive measurements, events with pions at extreme backward angles also require 
detection of recoil nucleons at extreme forward angles and at kinetic energies mostly 
in the punch-through regime. Such events have complicated detection efficiencies so that for 
this angular range, the inclusive analysis is more reliable than the result from the sum of 
the exclusive cross sections.

The situation for extreme pion forward angles is different. Systematic effects due to the sum 
trigger and the detection of the low-energy recoil nucleons are also important. 
However, there is also a physical reason for deviations because at extreme forward angles, coherent 
photoproduction of pions off the deuteron, the $\gamma d\rightarrow d\pi^0$ reaction, may contribute. 
Such events are included in the inclusive cross section, but not in the exclusive cross sections where 
identification of recoil protons or neutrons is required. Therefore, as observed, the cross section 
for the inclusive reaction can be larger. This is also related to the FSI effects. Nucleon-nucleon FSI, 
which, when it leads to a binding of the two nucleons in the final state, will shift strength from 
the exclusive quasi-free channels to the coherent reaction and thus deplete the exclusive reactions 
at forward angles. This makes the inclusive results interesting for testing models that 
investigate FSI effects.

The results for the exclusive, quasi-free cross sections with detection of coincident recoil nucleons
are summarized as angular distributions in Figs.~\ref{fig:dcsw_p} and \ref{fig:dcsw_n}, and as excitation 
functions for each angle bin in Fig.~\ref{fig:dcsw_e_np}. The deviation of the quasi-free proton 
data from the model results (see Figs.~\ref{fig:dcsw_p} and \ref{fig:dcsw_e_np}), which are only valid for 
free protons, is due to important FSI effects. The results from the SAID \cite{SAID,SAID_new}, 
MAID \cite{MAID,MAID_new}, and BnGa \cite{Anisovich_10} models for the free $\gamma p\rightarrow p\pi^0$ 
reaction are almost identical because all models have been fitted to the same large database for the 
production of $\pi^0$ mesons off free protons. 

The comparison of the present quasi-free proton data to the consistent model results for the free proton 
cross section (see Fig.~\ref{fig:dcsw_p}) demonstrates that the FSI effects vary in non-trivial ways. 
For example, they are much more important in the $W$ range between 1500 - 1550~MeV (i.e. in the second 
resonance region) than in the tail of the $\Delta$ resonance between 1450 - 1480~MeV. The different 
behavior of the data for the $p\pi^0$ and $n\pi^0$ final state, which is best seen in 
Fig.~\ref{fig:dcsw_e_np}, carries the physics information about the substantial isospin dependence of 
neutral pion production off protons and off neutrons. 

Figs.~\ref{fig:dcsw_corr_n} and \ref{fig:dcsw_e_ncorr} show the results for the neutron target 
corrected for FSI under the assumption that FSI is equal for quasi-free neutrons 
and protons (see Eq.~\ref{eq:fsi_corr}). Note that systematic uncertainties (in particular 
visible when comparing Fig.~\ref{fig:dcsw_e_np} and Fig.~\ref{fig:dcsw_e_ncorr}) are very different
from the quasi-free data for neutrons because several systematic effects (related to trigger 
thresholds, empty target, photon detection, invariant mass analysis, etc.) cancel in 
Eq.~\ref{eq:fsi_corr}. The 7\% overall normalization uncertainty also does not apply. The residual
uncertainty is dominated by the detection efficiency for recoil protons and neutrons (estimated 
from the comparison of inclusive data and the sum of exclusive cross sections), the systematic 
uncertainty of the world database for the cross section of the free $\gamma p\rightarrow p\pi^0$
reaction (which is negligible), and the folding of this cross section with the experimental 
resolution. Therefore, the systematic uncertainties for the extreme backward angles are much smaller
for the FSI corrected results (see Fig.~\ref{fig:dcsw_e_ncorr}) than for the originally measured
quasi-free neutron data (see Fig.~\ref{fig:dcsw_e_np}).

\begin{figure}[!thbp]
\centerline{\resizebox{0.50\textwidth}{!}{%
  \includegraphics{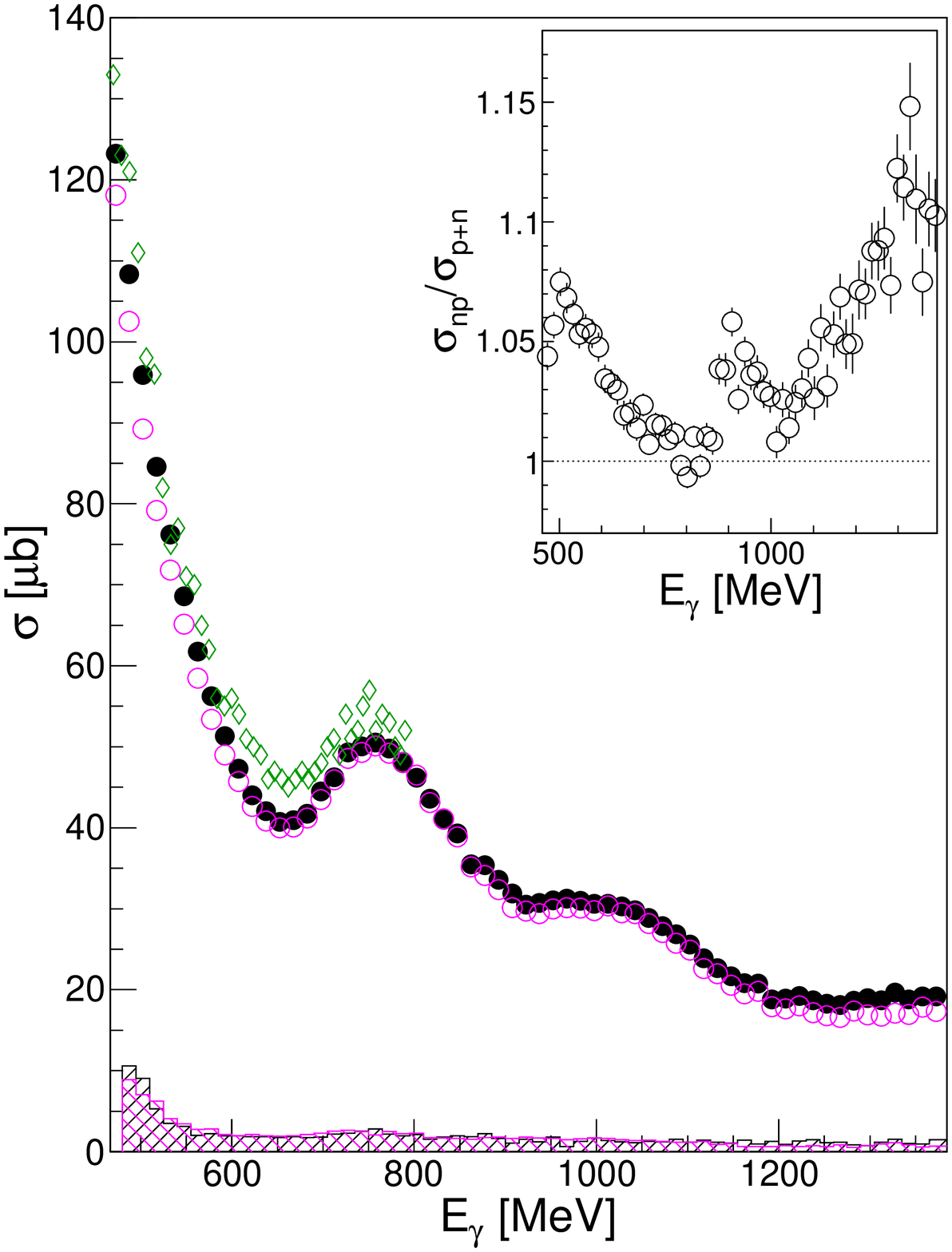}
}}
\caption{Total cross section as a function of the incident photon energy for quasi-free 
inclusive single $\pi^{0}$ photoproduction. Full (black) circles: quasi-free inclusive data, 
open (magenta) circles: sum of quasi-free proton and quasi-free neutron total cross section, 
open (green) diamonds: MAMI 99 quasi-free inclusive data \cite{Krusche_99}, 
hatched histograms: systematic errors. Insert: ratio of the inclusive cross section and
sum of the two exclusive cross sections.}
\label{fig:tcs_inc}       
\end{figure}

The data are compared in Figs.~\ref{fig:dcsw_p}-\ref{fig:dcsw_e_ncorr} to the most recent results 
from some reaction models (in particular those which provide results for the proton and 
neutron target). These are the BnGa coupled channel \cite{Anisovich_10,Anisovich_13},
the MAID~\cite{MAID,MAID_new}, and the SAID~\cite{SAID,SAID_new} analyses. Note that the references
refer only to the basic descriptions of the different analyses. The analyses evolve 
continuously and the most recent results are available on the respective websites \cite{pwaweb}.

In Figs.~\ref{fig:dcsw_p}-\ref{fig:dcsw_e_ncorr}, only the most recent results from the three 
models are compared to the data. They are partly different from the results shown in the preceding 
letter \cite{Dieterle_14} because in the meantime, a larger database has been included in the
fits of the BnGa and SAID analyses. This has not yet happened for the MAID model
and Figs.~\ref{fig:dcsw_corr_n} and \ref{fig:dcsw_e_ncorr} clearly show that this model is in 
poorer agreement with the experimental data. For the other models, some fine adjustments are
still necessary. 

Total cross sections $\sigma(W)$ have been derived from the angular distributions by fits of 
Legendre polynomials
\begin{equation}
\frac{d\sigma}{d\Omega} = 
\sum_{i=0}^{6} B_iP_i({\rm cos}(\Theta^{\star}_{\pi^0})) \ ,
\label{eq:legendre}
\end{equation}
using $\sigma(W) =4\pi B_0(W)$. The order of the expansion ($n=6$) was chosen such that the 
coefficient of this order was still significantly different from zero within statistical 
uncertainties. This analysis extrapolates the unmeasured differential cross sections at extreme 
forward angles. This effect is small below energies of $W\approx$1.6~GeV, but contributes 
more to the systematic uncertainty at larger $W$.

The total cross section $\sigma_{\rm incl}$ for the inclusive reaction is shown as a function of
$E_{\gamma}$ in Fig.~\ref{fig:tcs_inc}. The result from the inclusive analysis without any 
conditions on recoil nucleons and the sum of the exclusive cross sections $\sigma_p$ and $\sigma_n$ 
are compared. The agreement between the two data sets is excellent and demonstrates again
that systematic effects from the detection efficiency for the recoil nucleons must be small.
The insert in the figure shows the ratio of the results from these two analyses. Deviations are
within the 10\% range, but mostly smaller. The ratio is always above unity, which is reasonable 
because the sum of the exclusive cross sections excludes the contribution from the coherent 
$\gamma d\rightarrow d\pi^0$ reaction. At photon energies below 800~MeV, this effect alone can 
explain the deviations (see \cite{Krusche_99} for the relative contribution of the coherent 
reaction), at higher incident photon energies systematic uncertainties probably dominate. 

For photon energies below 800~MeV, the present data can be compared to the previous results from
\cite{Krusche_99}. They agree within their systematic uncertainties (typical deviations are of the
order of 10\%, the overall normalization of both data sets is $\approx$7\%, additional uncertainties
from analysis cuts etc. are $\approx$5\%). 

\begin{figure}[!thbp]
\centerline{\resizebox{0.5\textwidth}{!}{%
  \includegraphics{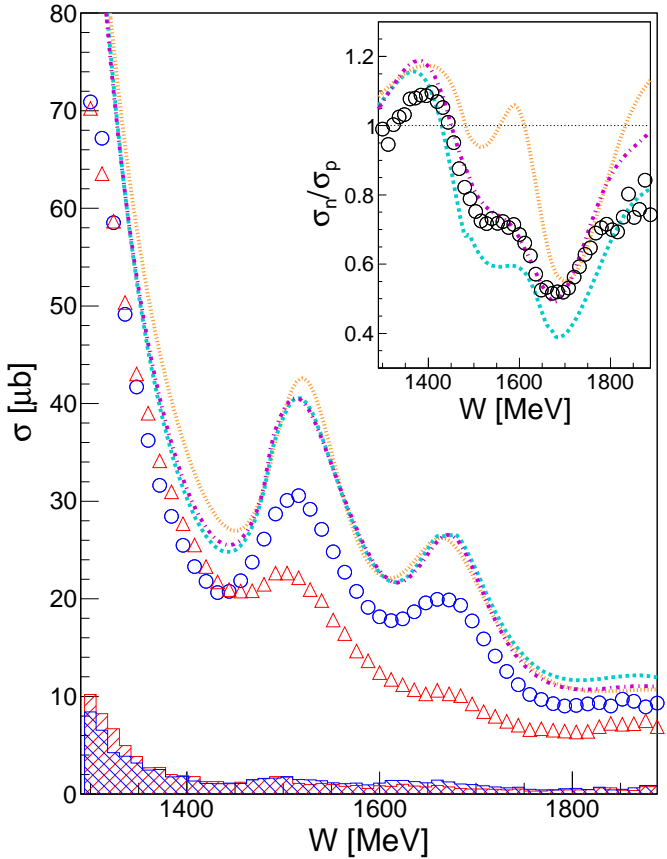}  
}}
\caption{Total cross section as a function of the final state invariant mass for exclusive 
single $\pi^{0}$ photoproduction off the quasi-free proton (open blue circles) and the 
quasi-free neutron (open red triangles). Dashed cyan line: SAID, dotted orange line: MAID, 
dash-dotted magenta line: BnGa. The insert shows the ratio of 
the quasi-free neutron to the quasi-free proton (open black circles).}
\label{fig:tcsw}       
\end{figure}

The total cross sections for the quasi-free reactions $\gamma d\rightarrow p(n)\pi^0$ and 
$\gamma d\rightarrow n(p)\pi^0$ (spectator nucleons in parentheses) are shown in Fig.~\ref{fig:tcsw}.
The results are compared to the predictions of the BnGa, MAID, and SAID analyses for the
free proton target. These predictions are similar, constrained by the same, large database of
the free $\gamma p\rightarrow p\pi^0$ reaction.  The figure demonstrates the substantial FSI effect 
on the quasi-free reaction even when nucleons are only bound in the lightest nucleus, the deuteron. 
In the maxima of the second resonance bump, this effect is on the order of 37\% and in the third 
resonance bump it is still around 30\%. 

In addition, the figure shows that the second and, even more so, the third resonance bumps are much 
less pronounced for quasi-free neutrons than for protons, while, due to the dominant reaction mechanism, 
these two cross sections are quite similar in the tail of the $\Delta$ resonance, as expected. 
This result sheds some new light on the suppression of the second and third resonance bump in the 
total photoabsorption on the deuteron compared to the free proton target \cite{Krusche_11}. 
Obviously, both mechanisms mentioned in the introduction play a role:
The quasi-free reaction on protons is damped compared to the free proton due to FSI effects, 
in particular in the maxima of the resonance peaks. Furthermore, both resonance peaks are much 
less pronounced for the quasi-free neutron than for the proton. This is due to the isospin structure 
of the excitation of the nucleon resonances involved. The insert in the figure shows the ratio of 
the total neutron and proton cross sections compared to model predictions. The SAID and 
BnGa analyses are in fair agreement with the measurements, but the MAID analysis overestimates the 
contribution of the $N(1525)3/2^-$ resonance for the neutron.

\begin{figure}[!htbp]
\centerline{\resizebox{0.5\textwidth}{!}{%
  \includegraphics{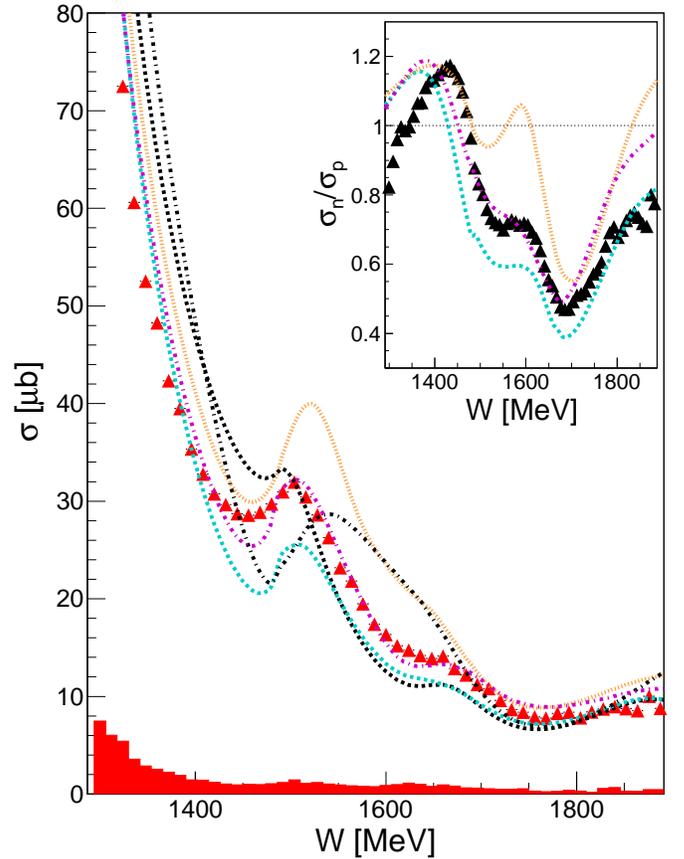}    
}}
\caption{Full red triangles: Total cross section as a function of the final state invariant mass 
for the free neutron (quasi-free neutron data corrected for FSI effects). 
Dashed cyan line: SAID, dotted orange line: MAID, dash-dotted 
magenta line: BnGa. The black dashed and dash-dotted lines show the results of the SAID and BnGa
analysis previous to the results from the present work and \cite{Dieterle_17}. The insert 
shows the ratio of the free neutron to the SAID proton (full black triangles).}
\label{fig:tcsw_n}       
\end{figure}

The results for the total cross section for $\gamma n\rightarrow n\pi^0$ (i.e. the quasi-free
$\gamma d\rightarrow \pi^0 n(p)$ data after removing effects from Fermi motion and with FSI corrections) 
are compared to model predictions in Fig.~\ref{fig:tcsw_n}. The experimental data are slightly 
changed with respect to the results shown in \cite{Dieterle_14} due to an improved 
treatment of the experimental resolution in the FSI correction. 

The results from the SAID and BnGa analyses, prior to the present experimental results and prior
to the data from Ref.~\cite{Dieterle_17} for the helicity dependence of the reaction, are also shown. 
They highlight the impact of the new quasi-free neutron data. Closest to the experimental results 
is the most recent fit of the BnGa model (note the large change of the results from this model 
compared to the previous fit). Agreement is slightly worse with the SAID results which did 
not much change by the inclusion of the recent quasi-free data. The MAID analysis clearly 
needs to be updated with inclusion of the recent quasi-free data. 

The experimental results for the $\sigma_n/\sigma_p$ ratio given in Figs.~\ref{fig:tcsw} and 
\ref{fig:tcsw_n} are quite similar. The values in Fig.~\ref{fig:tcsw} were directly obtained 
as a ratio of the measured total quasi-free cross sections $\sigma_n^{qf}/\sigma_p^{qf}$. 
The results in Fig.~\ref{fig:tcsw_n} represent the ratio $\sigma_n^f/\sigma_p^f$. 
Since $d\sigma_n^f/d\Omega$ was calculated from $d\sigma_n^{qf}/d\Omega$ by application of 
the FSI correction factors $<d\sigma_p^f>/d\sigma_p^{qf}$ (see Sec.~\ref{sec:FSI}), the correction 
cancels as long as it is independent on the polar angle $\theta_{\pi}^{\star}$ (which it almost is).

\begin{figure}[!thbp]
\centerline{\resizebox{0.5\textwidth}{!}{%
  \includegraphics{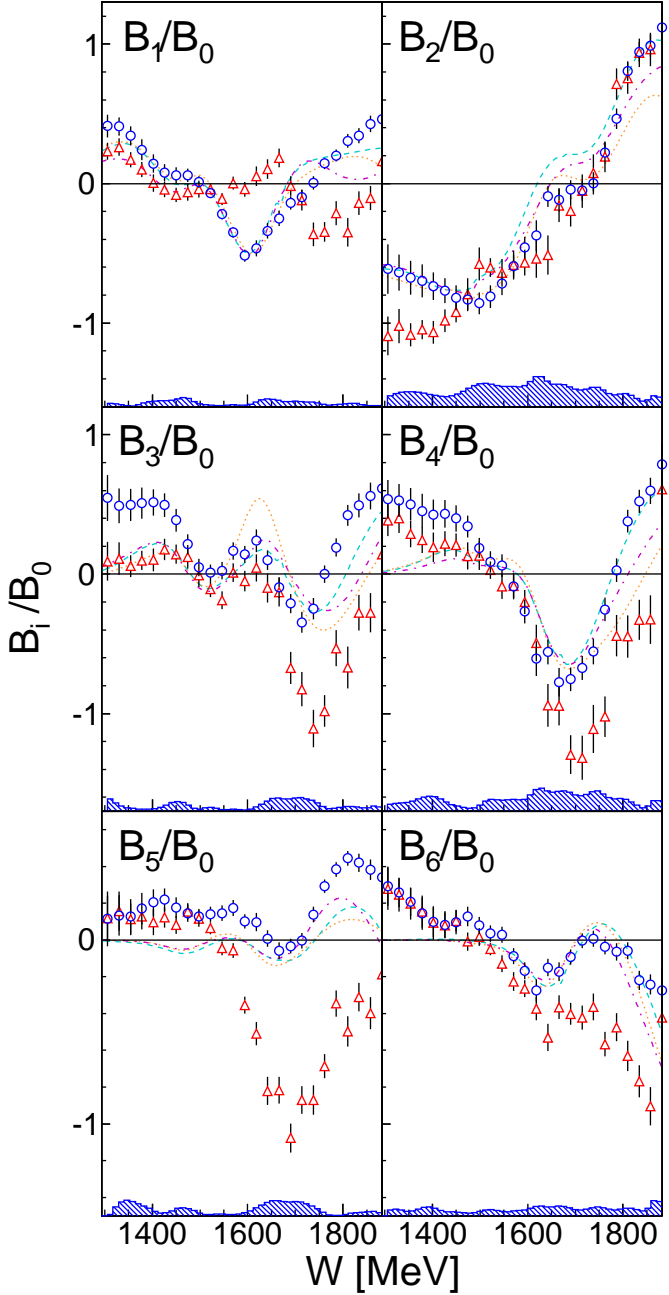}  
}}
\caption{Normalized Legendre coefficients as a function of the final state invariant mass for 
exclusive single $\pi^{0}$ photoproduction off the quasi-free proton (open blue circles) and 
the quasi-free neutron (open red triangles). Hatched histograms: systematic uncertainties of the 
quasi-free proton. Dashed cyan curve: SAID, dotted orange curve: MAID, 
dash-dotted magenta curve: BnGa.
}
\label{fig:legw}       
\end{figure}

\begin{figure}[!thbp]
\centerline{\resizebox{0.5\textwidth}{!}{%
  \includegraphics{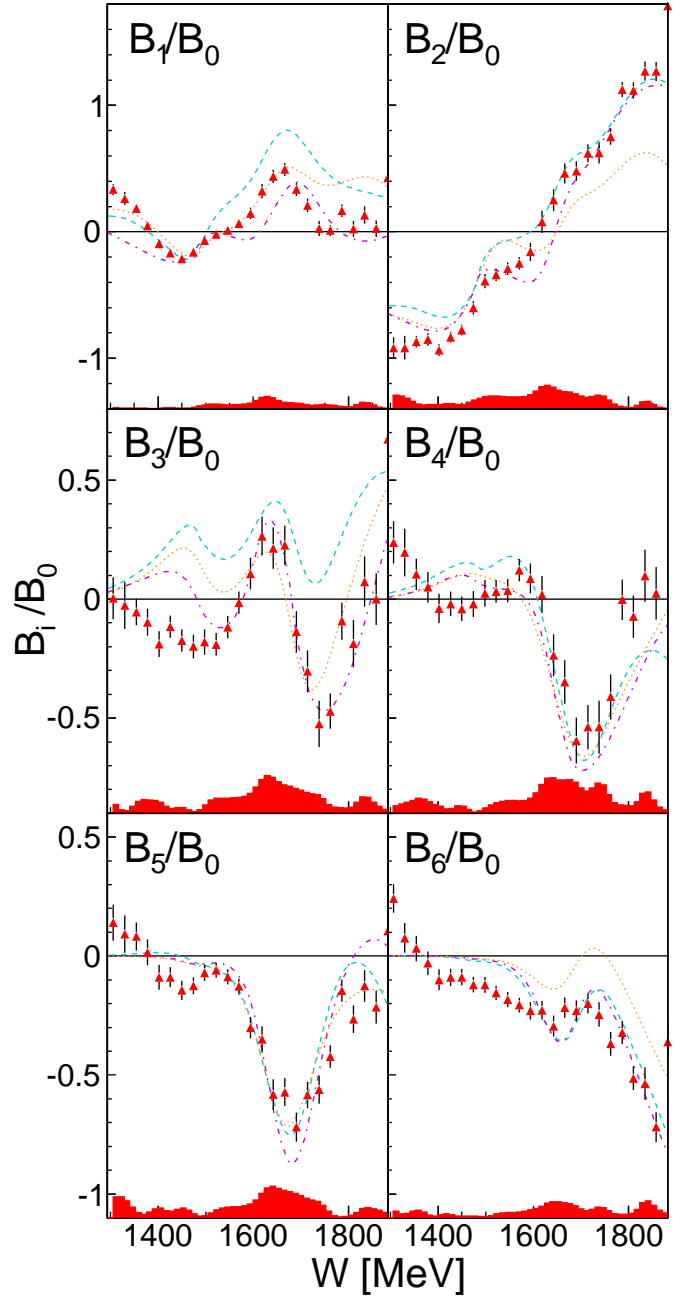}  
}}
\caption{Full red triangles: normalized Legendre coefficients as a function of the final state 
invariant mass for exclusive single $\pi^{0}$ photoproduction off the free neutron (quasi-free
data corrected for FSI effects). 
Solid histograms: systematic uncertainties, dashed cyan curve: SAID, dotted orange curve: MAID, 
dash-dotted magenta curve: BnGa.
}
\label{fig:legw_n}       
\end{figure}

The behavior of the angular distributions is reflected in the coefficients of the Legendre 
polynomials (Eq.~\ref{eq:legendre}) fitted to the experimental data. They are shown in 
Fig.~\ref{fig:legw} for the quasi-free data and in Fig.~\ref{fig:legw_n} for the extracted
free neutron data. All coefficients are normalized to the leading $B_0$, which is proportional
to the total cross section. Model results from BnGa, MAID, and SAID for the free proton are 
compared to the data in Fig.~\ref{fig:legw}, and those for the free neutron from the same analyses 
are shown in Fig.~\ref{fig:legw_n}. All model results were obtained by fits of the angular 
distributions with Eq.~\ref{eq:legendre} exactly as in the treatment of the experimental data.
Fig.~\ref{fig:legw} highlights the differences between
the $\gamma p\rightarrow p\pi^0$ and $\gamma n\rightarrow n\pi^0$ reactions for higher partial
waves, which usually don't leave large signals in the total cross section. In particular, around
invariant masses of 1.7~GeV - in the third resonance region - large signals are seen in the 
$B_3$ and $B_5$ coefficients for the neutron target. 

When such proton/neutron differences are due to resonance excitations, only $N^{\star}$ states 
can be responsible since electromagnetic $\Delta$ excitations are not isospin dependent. 
It was already emphasized in the preceding letter \cite{Dieterle_14} that, for example in the 
BnGa model, a refit to the previously existing database {\it and} the new neutron data mainly 
modified the resonant isospin $I=1/2$ partial waves and non-resonant backgrounds. The $I=3/2$ 
partial waves were much more stable because they are better constrained by the data for the 
free $\gamma p\rightarrow p\pi^0$ reaction. 

In the energy region around $W$=1.7~GeV, two $N^{\star}$ resonances with spin $J=5/2$ contribute, 
the $N(1675)5/2^-$ ($D_{15}$ partial wave) and the $N(1680)5/2^+$ ($F_{15}$). According to RPP
\cite{PDG16}, the $F_{15}$ has a much larger electromagnetic coupling to the proton and is 
responsible for a large fraction of the third resonance bump for the proton. The $D_{15}$ is one
of the few states which couple more strongly to the neutron. Its influence on the angular
distributions seems to be well reproduced by the BnGa and MAID model results, but significant
deviations are observed for the $B_{3}$ coefficient in this energy range for SAID 
(see Fig.~\ref{fig:legw_n}).

In Fig.~\ref{fig:legw_n}, the Legendre coefficients of the free $\gamma n\rightarrow n\pi^0$ 
reaction (constructed from the FSI corrected quasi-free neutron data) are compared to the reaction 
model results. A comparison of the quasi-free (Fig.~\ref{fig:legw}) and `free' (Fig.~\ref{fig:legw_n}) 
neutron data does not show much difference (the largest for the $B_3$ coefficient). This is again 
due to the fact that FSI seems mainly to act on the absolute scale of the cross sections 
(which is removed by the renormalization to the $B_0$ coefficient), but not so much on the shape 
of the angular distributions. The comparison to the model predictions does not allow a clear 
conclusion. Although on average, the MAID analysis agrees less well with the total cross section 
than the SAID results, some features, such as the behavior of the $B_3$ coefficient at high energies, 
are better reproduced by MAID than by SAID. Altogether, all reaction models will need readjustment
to accommodate the new neutron measurements.

\section{Summary and Conclusions}
Photoproduction of $\pi^0$ mesons from the deuteron has been measured in a high statistics experiment
with the Crystal Ball/TAPS detector at the electron accelerator MAMI in Mainz for incident photon
energies between 0.45~GeV and 1.4~GeV, corresponding approximately to cm energies in the photon-nucleon
system of 1.3~GeV to 1.875 GeV. Angular distributions were obtained in bins of
cos$(\theta^{\star}_{\pi^0})=0.1$ and only the extreme forward bin from 0.9 - 1.0 was not covered.
Data have been analyzed for the inclusive reaction $\gamma d\rightarrow X\pi^0$, where $X$ is either 
a neutron-proton pair or a deuteron. The reaction was identified by detection of the $\pi^0$ mesons 
and kinematic cuts excluding production of further mesons. Also analyzed were the exclusive reactions 
$\gamma d\rightarrow p\pi^0 (n)$ and $\gamma d\rightarrow n\pi^0 (p)$ in coincidence with recoil
protons or recoil neutrons where the nucleons in parentheses are undetected spectators.  

A comparison of the results from the inclusive reaction $\sigma_{\rm incl}$ to the sum of the exclusive
reactions $\sigma_p$, $\sigma_n$, sets stringent limits on systematic uncertainties of the detection
of recoil nucleons because $\sigma_{\rm incl}$ is completely independent of such effects. The inclusive 
data are of interest for the investigation of FSI effects because all event classes with production of one 
$\pi^0$ and no further meson are included without discrimination against different baryonic final states.

The most interesting experimental information comes from the investigation of the 
$\gamma n\rightarrow n\pi^0$ reaction. The present results represent the first comprehensive data set for
this reaction. The comparison to proton data demonstrates clearly the large isospin dependence 
of this reaction. The comparison to model results and PWA shows that analyses based only on data from
the other three isospin channels (the final states $p\pi^0$, $n\pi^+$, $p\pi^-$) are not sufficiently 
constrained. This was somehow expected because the model predictions disagreed significantly among
themselves. But it was also demonstrated, by the refit of one model, that the present and the 
previous data from other isospin channels can be accomodated in the same fit when the critical partial
waves (in particular those from excitations of $N^{\star}$ resonances and non-resonant backgrounds)
are properly adjusted. 

These results are not completely model independent.
Originally, the quasi-free $\gamma d\rightarrow n\pi^0(p)$ reaction was measured with a detected
`participant' neutron and an undetected `spectator' proton. The effective invariant mass $W$
of the intermediate state of the photon and the participant nucleon depends on nuclear Fermi motion.
This effect was removed by using the invariant mass $W$ derived from the detected pion and the
final-state participant nucleon. The resolution obtained for $W$, reconstructed this way, depends
on the detector resolution of the four momenta of the particles, rather than on the much
better resolution of the momenta of the degraded electrons in the tagging spectrometer.

Effects from nuclear FSI have been corrected under the assumption that it is equal for participant 
protons and neutrons. The ratio of free ($\gamma p\rightarrow p\pi^0$) and quasi-free 
($\gamma d\rightarrow p\pi^0 (n)$) proton production cross sections was used to correct the quasi-free 
neutron data. The available results from modeling FSI effects \cite{Tarasov_16} support the assumption 
that, for the angular range covered by the experimental data, they are similar for participant 
protons and neutrons. However, these results \cite{Tarasov_16} are not in quantitative agreement with 
the experimental proton data so that further refinements of the FSI modeling are required before it 
can be used for reliable FSI corrections of quasi-free neutron data.

It is obvious from the comparison of the most recent reaction model analyses from BnGa, MAID, 
and SAID \cite{Anisovich_13,MAID_new,SAID_new} to the present neutron data that these analyses still 
need refinements, which will help to establish a more solid database for electromagnetic excitations 
of neutron $N^{\star}$ resonances.  

\begin{acknowledgments}
We wish to acknowledge the outstanding support of the accelerator group and operators of MAMI. 
This work was supported by Schweizerischer Nationalfonds (200020-156983, 132799, 121781, 117601), 
Deutsche For\-schungs\-ge\-mein\-schaft (SFB 443, SFB 1044, SFB/TR16), the INFN-Italy, 
the European Community-Research Infrastructure Activity under FP7 programme (Hadron Physics, 
grant agreement No. 227431), 
the UK Science and Technology Facilities Council (ST/J000175/1, ST/G008604/1, ST/G008582/1,ST/J00006X/1, and 
ST/L00478X/1), 
the Natural Sciences and Engineering Research Council (NSERC, FRN: SAPPJ-2015-00023), Canada. This material 
is based upon work also supported by the U.S. Department of Energy, Office of Science, Office of Nuclear 
Physics Research Division, under Award Numbers DE-FG02-99-ER41110, DE-FG02-88ER40415, and DE-FG02-01-ER41194 
and by the National Science Foundation, under Grant Nos. PHY-1039130 and IIA-1358175.
\end{acknowledgments}

\end{document}